\documentclass[final,authoryear,5p,times,twocolumn]{elsarticle}

\usepackage{graphicx}

\usepackage{amssymb}
\usepackage{amsmath}
\usepackage{enumerate}
\usepackage{enumitem}
\usepackage{times}
\usepackage{jtb}
\usepackage{mathtools}

\DeclareMathAlphabet{\bi}{OML}{cmm}{b}{it}

\newcommand{\vv}[1]{\mathbf{#1}}

\journal{Journal of Theoretical Biology}

\begin{document}

\begin{frontmatter}



\title{Stochastic competitive exclusion leads to a cascade of species extinctions}


\author[upm,ceab]{Jos\'e A.\ Capit\'an\corref{cor}}
\ead{ja.capitan@upm.es}
\cortext[cor]{Corresponding author.} 
\address[upm]{Department of Applied Mathematics, Technical University of Madrid, Madrid, Spain}
\author[uam]{Sara Cuenda}
\address[uam]{Departamento de An\'alisis Econ\'omico: Econom{\'\i}a Cuantitativa, Universidad 
Aut\'onoma de Madrid, Madrid, Spain}
\ead{sara.cuenda@uam.es}
\author[ceab]{David Alonso}
\address[ceab]{Center for Advanced Studies (CEAB-CSIC), Blanes, Catalunya, Spain}
\ead{dalonso@ceab.csic.es}

\begin{abstract}
Community ecology has traditionally relied on the competitive exclusion principle, 
a piece of common wisdom in conceptual frameworks developed to describe species 
assemblages. Key concepts in community ecology, such as limiting similarity and niche 
partitioning, are based on competitive exclusion. However, this classical paradigm in 
ecology relies on implications derived from simple, deterministic models. Here we show 
how the predictions of a symmetric, deterministic model about the way extinctions proceed 
can be utterly different from the results derived from the same model when ecological drift 
(demographic stochasticity) is explicitly considered. Using analytical approximations to 
the steady-state conditional probabilities for assemblages with two and three species,
we demonstrate that stochastic competitive exclusion leads to a cascade of extinctions,
whereas the symmetric, deterministic model predicts a multiple collapse of species.
To test the robustness of our results, we have studied the effect of environmental stochasticity 
and relaxed the species symmetry assumption. Our conclusions highlight the crucial role of 
stochasticity when deriving reliable theoretical predictions for species community assembly.
\end{abstract}


\end{frontmatter}


\section{Introduction}
\label{sec:intro}

Ecological communities are shaped from the complex interplay of 
four fundamental processes~\citep{vellend:2010}: selection, in the form of species
interactions that favor certain species against others; speciation, leading to the
appearance of new species, better adapted to the environment; dispersal, which
permits spatial propagation of individuals; and ecological drift, a demographic 
variability in species population numbers due to the stochastic processes that take 
place. Ecological drift, in particular, has a prevalent role in modern theoretical 
frameworks in community ecology~\citep{black:2012}. Accordingly, current approaches 
reveal the need for process-based, stochastic models that help to understand how 
ecological communities are assembled and their interaction with environmental 
factors~\citep{wisz:2013}. 

Classical community ecology, however, has mainly relied on deterministic 
community models (see~\cite{roughgarden:1979} and references therein), most of
them based on Lotka-Volterra dynamics, although alternatives have been 
proposed~\citep{schoener:1976}. There is a long-standing research focus on community 
assembly models, in which communities are built up through species invasions,
and most of them rely on deterministic approaches~\citep{post:1983,law:1993,law:1996,capitan:2009,capitan:2011a,capitan:2011b}.
On the other side, there have been strong theoretical efforts to describe community
assemblages in stochastic terms~\citep{hubbell:2001,alonso:2008,rosindell:2011}. 
In certain situations, the results and conclusions derived from deterministic models 
have been shown to be quite different in the presence of 
stochasticity~\citep{bolker:1995,alonso:2007,haegeman:2011,bonachela:2012,wang:2012}.

One of the contexts where the differences between deterministic and stochastic
approaches become apparent is related to theoretical formulations of the competitive 
exclusion principle~\citep{volterra:1926,gause:1934,hardin:1960}. This principle
constitutes a fundamental pillar of community ecology and belongs to the
traditional body of ecological theory. It provides a useful theoretical framework to 
explore how complex species assemblages persist over time. Important
concepts such as adaptation to shared niches~\citep{roughgarden:1979}, 
species limiting similarity~\citep{macarthur:1967a,roughgarden:1974} or niche 
partitioning~\citep{pielou:1972,schoener:1974} all are immediate derivations of 
the principle. Classical approaches predict the maximum degree of species similarity 
that permit species stable coexistence~\citep{macarthur:1969,macarthur:1970}.
However, theoretical predictions for limiting similarity often rely on deterministic
community models (see~\cite{macarthur:1968,levin:1970,haigh:1972,chesson:1990} 
and~\ref{sec:CEP} for a discussion on competitive exclusion based on deterministic 
approaches), and the relevance of stochasticity, in the form of ecological drift, to 
species coexistence has remained almost unexplored (with the exception of~\cite{turelli:1980}). 
The relationship between limiting similarity and environmental stochasticity has been
studied more thoroughly~\citep{may:1972b,turelli:1978,turelli:1981}.

Recently, we focused on the influence of ecological drift on the similarity of coexisting 
species via the competitive exclusion principle~\citep{capitan:2015}. In that contribution 
we showed that, in the presence of ecological drift, the maximum degree of similarity 
that ensures stable coexistence can be significantly lowered when compared to 
the corresponding limits to similarity derived from deterministic models. 
If similarity is interpreted in terms of an interspecific competitive 
overlap~\citep{macarthur:1967a,roughgarden:1974}, stochasticity 
displaces the deterministic threshold towards lower values of the competitive 
overlap~\citep{capitan:2015}. Thus, when stochasticity is considered, the extinction 
phenomena caused by competitive exclusion takes place at lower values of the
competitive overlap (i.e., species have to be more dissimilar to stably coexist in
the presence of ecological drift).

Ecological drift becomes a key process determining species coexistence in aspects
other than the maximum similarity of co-occurring species. Beyond a more restrictive 
threshold in competition induced by ecological drift (which was the main result 
of~\cite{capitan:2015}), we here analyze the influence of demographic stochasticity on 
the extinction mechanism itself, which in principle can lead to either sequential 
or grouped extinctions as competition strength increases. For that purpose, 
we considered a deterministic, Lotka-Volterra model and its stochastic counterpart, both 
of which treat species interactions symmetrically. Whereas the deterministic model predicts 
the multiple extinction of all the species in the community but one as competition
crosses over a certain threshold, in the presence of demographic stochasticity extinctions 
proceed progressively, in the form of a cascade, as competition increases. 
The only difference between both approaches is the
explicit consideration of ecological drift in the dynamics. In order to derive our 
conclusions, we developed convenient analytical approximations to the 
steady-state configurations of the stochastic system for simple species assemblages 
formed by two or three species. Such approximations help us to partition the set of feasible 
population numbers into regions associated to coexistence, or the extinction of one, 
two, or three species. The steady-state probabilities, when aggregated over those 
regions, unveil the extinction cascade phenomenon. Our main result reveals overlapping 
windows in competitive strength, at low values related to configurations where the 
coexistence of three species is the most 
probable state, intermediate ranges where it is more likely to observe two-species 
assemblages, and large competition values for which the most probable state is 
formed by one species or none. We also studied the transition to the deterministic 
model when demographic stochasticity tends to zero, and our results reveal an
abrupt transition to situations compatible with small stochasticity.

To test the robustness of our conclusions, we replaced demographic stochasticity by
environmental stochasticity and confirmed that, although the extinction phenomena are qualitatively
different, the extinction cascade persists. We also relaxed the assumption of symmetry
to assess the effect of stochasticity on deterministic models that not only predict 
multiple extinctions, as in the fully symmetric scenario, but also lead to both 
progressive and grouped extinctions for fixed competitive
strengths. When stochasticity comes into play, however, the stochastic 
cascade persists and the expected extinction sequence is qualitatively different from
its deterministic counterpart. Thus, the predictions of both models are 
significantly different in generic, non-symmetric scenarios for species interactions.

The paper is organized as follows: in Section~\ref{sec:stoch} we describe both 
the deterministic and the stochastic frameworks, the latter based on the formulation 
of~\cite{haegeman:2011}, and show that the deterministic, symmetric model 
predicts a multiple species extinction. In Section~\ref{sec:results} we start 
by presenting the analytical approximations for a 
two-species stochastic community model, and we then extend the procedure to a 
three-species community. These approximations help us to obtain analytical formulae 
for the critical points of the steady-state, joint probability distribution of the community. 
Formulae for saddle points are then used to properly define aggregated 
probabilities of coexistence, or one-, two-, and three-species extinction, which reveal 
themselves the sequential decline of species driven by ecological drift. After studying
the small stochasticity limit and testing the robustness of our results, we conclude
the paper with several implications and prospects (Section ~\ref{sec:discussion}).

\section{Model description}
\label{sec:stoch}

For the sake of simplicity, in this contribution we will focus on the symmetric 
version of the deterministic Lotka-Volterra competitive dynamics (see~\ref{sec:CEP}),
\begin{equation}\label{eq:dyn}
\dot{x}_i=rx_i\left(1-\frac{x_i+\rho\sum_{j\ne i}x_j}{K}\right),
\qquad i=1,\dots,S,
\end{equation}
where $x_i$ stands for the population density of species $i$ (space is implicitly assumed) 
and model parameters are uniform and species-independent. Here $r$ stands
for an intrinsic, species-independent growth rate, $\rho$ measures interspecific competition, 
$K$ represents a carrying capacity, and $S$ is the species richness of the 
community. The dynamics has an interior equilibrium point, 
$\vv{\hat{x}}=(\hat{x},\dots,\hat{x})$, where $\hat{x}=K/(1-\rho+\rho S)$,
which is globally stable if and only if $\rho<1$~\citep{hofbauer:1998,capitan:2015}.
In the symmetric scenario, the competitive exclusion principle adopts a very simple 
formulation (see~\ref{sec:CEP} for further details on the general, non-symmetric 
case). A complete stability analysis of the boundary equilibrium points shows that, 
for $\rho>1$, all the species become extinct except for just one of 
them (see~\ref{sec:appA}). As a consequence, competitive exclusion in the 
symmetric, deterministic model implies the joint extinction of $S-1$ 
species. 

We now explicitly incorporate ecological drift (demographic stochasticity) in the symmetric 
scenario in order to show that species are sequentially displaced in the presence 
of stochasticity due to competitive exclusion, following a cascade of extinctions,
as competition strengthens. 
A standard way to extend deterministic models to incorporate ecological drift
is deeply described in~\cite{haegeman:2011}. The state of the system is described 
by the vector of population numbers $n_i$ at time $t$, $\vv{n}(t)=(n_1(t),\dots,n_S(t))$.
Contrary to the deterministic case, which focuses on population densities 
$x_i=n_i/\Omega$, $\Omega$ being a meaningful measure (area, volume) of the 
system size, here discrete population numbers are considered. The elementary 
processes that
define the stochastic dynamics (local births and deaths, immigration, and competition)
are characterized by probability rates that, in the deterministic limit, yield the
Lotka-Volterra equations~\eqref{eq:dyn}. As in~\cite{haegeman:2011}, we choose
the following probability rates to model elementary processes:
\begin{enumerate}
\item Local births (deaths) of species $i$ occur at a density-independent rate 
$r^+n_i$ ($r^-n_i$). We adopt the notation $r=r^+-r^-$ to represent the net growth
rate in the absence of competitors.
\item Immigration of a new individual of species $i$ takes place at a rate $\mu$.
Although the deterministic model~\eqref{eq:dyn} does not include immigration,
dispersal is an important process driving community assembly~\citep{vellend:2010}.
In addition, immigration is key for the stochastic process to reach a non-trivial 
steady-state. We consider here the low-immigration regime, in which the deterministic 
limit is expected to recover results close to those yielded by Eq.~\eqref{eq:dyn},
see~\cite{capitan:2015}.
\item Intraspecific competition occurs at a density-dependent rate $rn_i^2/K$, 
where $K$ represents a species carrying capacity.
\item Interspecific competition between species $i$ and $j$ ($i\ne j$) takes 
place at a probability $r\rho n_in_j/K$ per unit time (it is also a density-dependent 
rate).
\end{enumerate}

Population vectors $\vv{n}(t)$ belong to
the configuration space $\mathbb{N}^S$, where $\mathbb{N}=\{0,1,2,\dots\}$. 
The elementary processes listed above define a birth-death-immigration stochastic 
process in continuous time, and the probability $P(\vv{n},t)$ of finding a 
population vector $\vv{n}(t)$ at time $t$ is determined by the master equation,
\begin{multline}\label{eq:master}
\frac{\partial P(\vv{n},t)}{\partial t}=\sum_{i=1}^S\left\{
q_i^+(\vv{n}-\vv{e}_i)P(\vv{n}-\vv{e}_i,t)\right.\\
\left.+q_i^-(\vv{n}+\vv{e}_i)P(\vv{n}+\vv{e}_i,t)-
[q_i^+(\vv{n})+q_i^-(\vv{n})]P(\vv{n},t)\right\}.
\end{multline}
Here $q_i^+(\vv{n})=r^+n_i+\mu$ is the overall birth rate for species $i$,
$q_i^-(\vv{n})=r^-n_i+rn_i(n_i+\rho\sum_{j\ne i}n_j)/K$ is the overall death rate,
and $\vv{e}_i=(0,\dots,1,\dots,0)$ is the $i$-th vector of the canonical
basis of $\mathbb{R}^S$. Notice the correspondence of these rates with 
the terms arising in the deterministic model~\eqref{eq:dyn} for $\mu=0$.
The steady-state probability distribution is obtained by solving
the coupled recurrence equation given by the condition 
$\partial P(\vv{n},t)/\partial t=0$ (see~\ref{sec:appB} for details).

Our approach develops analytical approximations for the critical points
of the joint probability function, not for the probability itself. For $S=1$
the stationary state of the birth-death-immigration
model can be exactly solved in terms of hypergeometric
functions~\citep{haegeman:2011}. For $S>1$, in the absence of competition ($\rho=0$)
populations are uncorrelated, and the joint probability distribution factors as a 
product of marginal probabilities, which reduces this case to a one-dimensional problem. 
The fully neutral model ($\rho=1$) for $S>1$ can be 
solved as well~\citep{haegeman:2011}. We cannot find analytically the 
steady-state distribution for $S\ge 2$ and $\rho>0$, though. \cite{haegeman:2011} 
devised approximations for this case, but we will follow here a different approach 
to find analytical formulae for the critical points of the steady-state distribution 
in the case of small-sized communities.

\section{Results}
\label{sec:results}

In~\cite{capitan:2015} we showed numerically that the steady-state distribution for 
two-species communities presents a maximum at an interior point of the configuration 
space, as well as two boundary maxima with population vectors of the form $(n,0)$ 
and $(0,n)$. This implies that the discrete probability distribution, when extended to 
be real-valued (by, for instance, cubic spline interpolation), must exhibit by continuity two 
saddle points located in between the coexistence maximum and the two boundary maxima. 
These saddle points can be used to conveniently partition the configuration space into 
regions associated to coexistence, the extinction of one species, or the extinction of two species. 
In this section we develop analytical approximations that help us to obtain good estimates 
for the critical points of the joint probability distribution. We use the $S=2$ case to illustrate
the technique. We then extend the method for three-species assemblages,
and use the approximated saddle points to partition the configuration space into regions 
for coexistence and for the three possible states where extinctions have occurred. The 
aggregation of the joint probability over those regions unveils the extinction cascade 
phenomenon.

\subsection{Critical points for two-species communities}
\label{sec:cp2}

To estimate the location of the critical points of the joint distribution $P(n_1,n_2)$
we use that the conditions $\partial P(n_1,n_2)/\partial n_1=0$ and
$\partial P(n_1,n_2)/\partial n_2=0$ are equivalent to
\begin{equation}
\begin{aligned}
\frac{\partial}{\partial n_1}P(n_1|n_2)=
\frac{\partial}{\partial n_1}\left(\frac{P(n_1,n_2)}{P(n_2)}\right)=0,\\
\frac{\partial}{\partial n_2}P(n_2|n_1)=
\frac{\partial}{\partial n_2}\left(\frac{P(n_1,n_2)}{P(n_1)}\right)=0,\\
\end{aligned}
\end{equation}
$P(n_1|n_2)$ being the probability that species 
$1$ has $n_1$ individuals conditioned to species $2$ having $n_2$ individuals. 
This means that critical points of the joint distribution $P(n_1, n_2)$ 
can also be obtained through conditional probabilities. 
By fixing $n_2$, we just need to evaluate the derivative of $P(n_1|n_2)$ along 
the $n_1$ direction. The same applies under the change $n_1\leftrightarrow n_2$.

\subsubsection{Approximated conditional probabilities}
We now approximate $P(n_2|n_1)$ by $T(n_2|n_1)$ as follows. For $S=2$, 
the steady-state distribution satisfies a two-term recurrence in population
numbers $n_1$ and $n_2$, namely
\begin{equation}\label{eq:eqs2}
\begin{aligned}
0&=q^+_1(n_1-1,n_2)P(n_1-1,n_2)+q^-_1(n_1+1,n_2)P(n_1+1,n_2)\\
&+q^+_2(n_1,n_2-1)P(n_1,n_2-1) +q^-_2(n_1,n_2+1)P(n_1,n_2+1)\\
&-\left[q^+_1(n_1,n_2)+q^-_1(n_1,n_2)\right]P(n_1,n_2)\\
&-\left[q^+_2(n_1,n_2)+q^-_2(n_1,n_2)\right]P(n_1,n_2).
\end{aligned}
\end{equation}
Notice that we are interested in approximating the conditional probability $P(n_2|n_1)$ 
where $n_1$ is fixed. Hence, in the approximation we ignore the terms in 
Eq.~\eqref{eq:eqs2} that involve 
variation of $n_1$, and assume that $n_1$ acts as a fixed parameter in the 
remaining terms. Thus, the approximated conditional probability $T(n_2|n_1)$ 
satisfies
\begin{equation}\label{eq:appconds2}
\begin{aligned}
0&=q^+_2(n_1,n_2-1)T(n_2-1|n_1)+q^-_2(n_1,n_2+1)T(n_2+1|n_1)\\
&-\left[q^+_2(n_1,n_2)+q^-_2(n_1,n_2)\right]T(n_2|n_1).
\end{aligned}
\end{equation}
This expression fulfills a detailed balance condition~\citep{karlin:1975}, which 
yields an approximate one-term recurrence formula in $n_2$,
\begin{equation}\label{eq:recconds2}
\begin{aligned}
T(n_2|n_1)&=\frac{q^+_2(n_1,n_2-1)}{q^-_2(n_1,n_2)}T(n_2-1|n_1)\\
&=\frac{K[\mu+r^+(n_2-1)]}{n_2[Kr^- +r(n_2+\rho n_1)]}T(n_2-1|n_1),
\end{aligned}
\end{equation}
and leads to an explicit solution in terms of hypergeometric functions,
as in~\cite{haegeman:2011}. A symmetric recurrence holds for $T(n_1|n_2)$.

\subsubsection{Analytical formulae for critical points}
Our next step is to approximate the partial derivative along the $n_1$ direction 
by the backwards discrete difference (comparable results are obtained with the 
forward difference),
\begin{equation}
\frac{\partial}{\partial n_1}P(n_1|n_2)\approx 
\frac{\partial}{\partial n_1}T(n_1|n_2)\approx T(n_1|n_2)-T(n_1-1|n_2).
\end{equation}
Similarly,
\begin{equation}
\frac{\partial}{\partial n_2}P(n_2|n_1)\approx T(n_2|n_1)-T(n_2-1|n_1).
\end{equation}
Eq.~\eqref{eq:recconds2} allows us to write the system for the critical 
points of the two-dimensional joint-probability surface,
$\partial P(n_1|n_2)/\partial n_1 = \partial P(n_2|n_1)/\partial n_2 = 0$,
as
\begin{equation}\label{eq:pcrits2}
\begin{aligned}
&n_1^2+\rho n_1n_2 -K n_1+K\alpha = 0,\\
&n_2^2+\rho n_1n_2 -K n_2+K\alpha = 0,
\end{aligned}
\end{equation}
where $\alpha=(r^+-\mu)/r$. Solving the quadratic system yields the following 
estimates for the interior critical points: $M_1=(m_+,m_+)$, which can be either
a local maximum or a saddle point depending on the value of $\rho$ (see below), 
and the local minimum $M_2=(m_-,m_-)$, where
\begin{equation}\label{eq:intmaxS2}
m_{\pm}=\frac{K}{2(1+\rho)}\left(1\pm\sqrt{1-\frac{4\alpha(1+\rho)}{K}}\right),
\end{equation}
and two (symmetrical) saddle points $Q_1=(s_+,s_-)$ and $Q_2=(s_-,s_+)$, where
\begin{equation}\label{eq:saddleS2}
s_{\pm}=\frac{K}{2}\left(1\pm\sqrt{1-\frac{4\alpha}{K(1-\rho)}}\right).
\end{equation}

\begin{figure*}[t!]
\begin{center}
\includegraphics[width=1\textwidth]{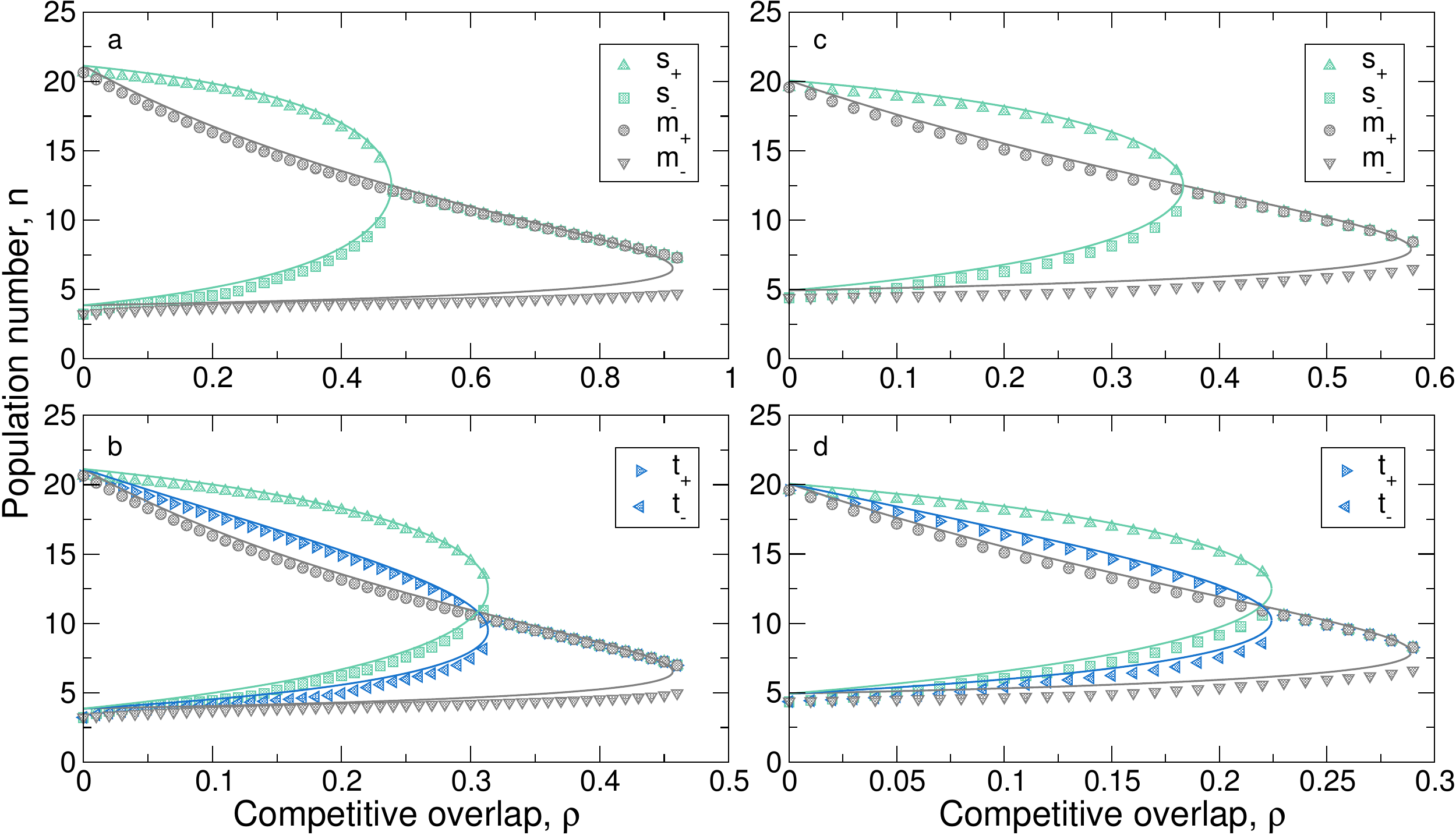}
\caption{\label{fig:cp}
(a) Coordinates of the critical points for $S=2$ as a function of $\rho$. 
They were calculated using the exact steady-state distribution 
(symbols) as well as the approximations given by 
Eqs.~\eqref{eq:intmaxS2} and~\eqref{eq:saddleS2} (lines). Once
the coexistence maximum $M_1=(m_+,m_+)$ and the two saddle points
$Q_i=(s_{\pm},s_{\mp})$ ($i=1,2$) coalesce, $M_1$ becomes a
saddle point (there is no longer a coexistence maximum). Model
parameters are $r^+=50$, $r^-=35$, $\mu=1$, $K=25$. (b) 
Coordinates of the critical points for a three-species community 
(symbols), compared with theoretical approximations~\eqref{eq:intmaxS3} 
and~\eqref{eq:saddleS3}. When the two saddle points coincide,
the former maximum becomes a saddle point. Parameter
values are the same as in panel (a). Panels (c)-(d) check the 
accuracy of our approximations for other set of parameter values,
namely $r^+=10$, $r^-=7.5$, $\mu=0.1$, and $K=25$. As before,
(c) represents the coordinates of the critical points for $S=2$
and (d) for $S=3$.
}
\end{center}
\end{figure*}

To test the goodness of these analytical approximations, critical points 
of the exact probability distribution are determined
numerically by extending the discrete distribution $P(n_1,n_2)$ to be a
real-valued function using cubic spline interpolation both in the $n_1$ and
$n_2$ directions. We then solve numerically the system $\partial
P(n_1,n_2)/\partial n_1 = \partial P(n_1,n_2)/\partial n_2 = 0$, using analytical
predictions as initial guesses for iterative root finding. The Hessian 
matrix decides whether a given point is a local maximum, a local minimum 
or a saddle point. We find that the critical points are in excellent 
agreement with the analytical formulae above (see Fig.~\ref{fig:cp}a, 
in which we plot the coordinates of each critical point calculated both 
analytically and numerically). 

For $\rho < \rho_c$, with
\begin{equation}
\rho_{\mathrm{c}}=1-\frac{4\alpha}{K}=1-\frac{4(r^+-\mu)}{Kr},
\end{equation}
the analysis of the Hessian matrix reveals that
$M_1$ is a (coexistence) local maximum. For $\rho=\rho_c$,
saddle points $Q_1$ and $Q_2$ coincide with $M_1$ and, 
for $\rho > \rho_c$, 
$M_1$ transforms into a saddle point and the former saddle points $Q_1$ and $Q_2$ 
(cf. Eq.~\eqref{eq:saddleS2}) no longer exist. Moreover, when $\rho>K/4\alpha-1$,
all interior critical points~\eqref{eq:intmaxS2} become complex and the only persistent
maxima are those located at the boundary. 

There are four critical points at the boundary, which can be
approximated using Eq.~\eqref{eq:recconds2} for $n_1=0$, resulting in two local
maxima [$(b_+,0)$, $(0,b_+)$] and two local minima [$(b_-,0)$, $(0,b_-)$],
where
\begin{equation}\label{eq:bdmaxS2}
b_{\pm}=\frac{K}{2}\left(1\pm\sqrt{1-\frac{4\alpha}{K}}\right).
\end{equation}
Observe that the non-zero coordinates of the local maxima (minima) coincide with
those of the interior maximum (minimum) for $\rho=0$.

\subsection{Critical points for three-species communities}
\label{sec:cp3}

The method proposed in the previous subsection can be fully extended to the 
case of a three-species community. Critical points are obtained by approximating 
conditional probabilities $P(n_1|n_2,n_3)$ and taking discrete derivatives
with respect to the first argument. As before, we consider the three-term
recurrence relation that fulfills the joint distribution $P(n_1,n_2,n_3)$ and
ignore the terms that involve variation in population numbers $n_2$ and
$n_3$. Under this approximation, the steady-state condition reduces to
\begin{equation}\label{eq:appconds3}
\begin{aligned}
0&=q^+_1(n_1-1,n_2,n_3)T(n_1-1|n_2,n_3)\\
&+q^-_1(n_1+1,n_2,n_3)T(n_1+1|n_2,n_3)\\
&-\left[q^+_1(n_1,n_2,n_3)+q^-_1(n_1,n_2,n_3)\right]T(n_1|n_2,n_3).
\end{aligned}
\end{equation}
Due to detailed balance, the approximate conditional probabilities 
$T(n_1|n_2,n_3)$ satisfy the one-term recurrence relation
\begin{equation}\label{eq:recs3p}
\begin{split}
\MoveEqLeft K\left[\mu+r^+(n_1-1)  \right]T(n_1-1|n_2,n_3)=\\
&n_1\left[ Kr^- +rn_1 + r\rho (n_2+n_3) \right]T(n_1|n_2,n_3).
\end{split}
\end{equation}

Repeating for $S=3$ the procedure devised to estimate the coordinates of
the critical points leads to the set of quadratic equations
\begin{equation}
\begin{aligned}
&n_1^2+\rho n_1(n_2+n_3) -K n_1+K\alpha = 0,\\
&n_2^2+\rho n_2(n_1+n_3) -K n_2+K\alpha = 0,\\
&n_3^2+\rho n_3(n_1+n_2) -K n_3+K\alpha = 0,
\end{aligned}
\end{equation}
which yields $8$ interior critical points, $6$ of which
are saddle points, and the remaining two points are a local minimum and, as in
$S=2$, a point that is a maximum or a saddle point depending on $\rho$.
Explicit expressions for their coordinates are given below. For the
sake of comparison we have also calculated critical points using the exact
joint distribution $P(n_1,n_2,n_3)$, see Fig.~\ref{fig:cp}b.

\begin{figure*}[t!]
\begin{center}
\includegraphics[width=1\textwidth]{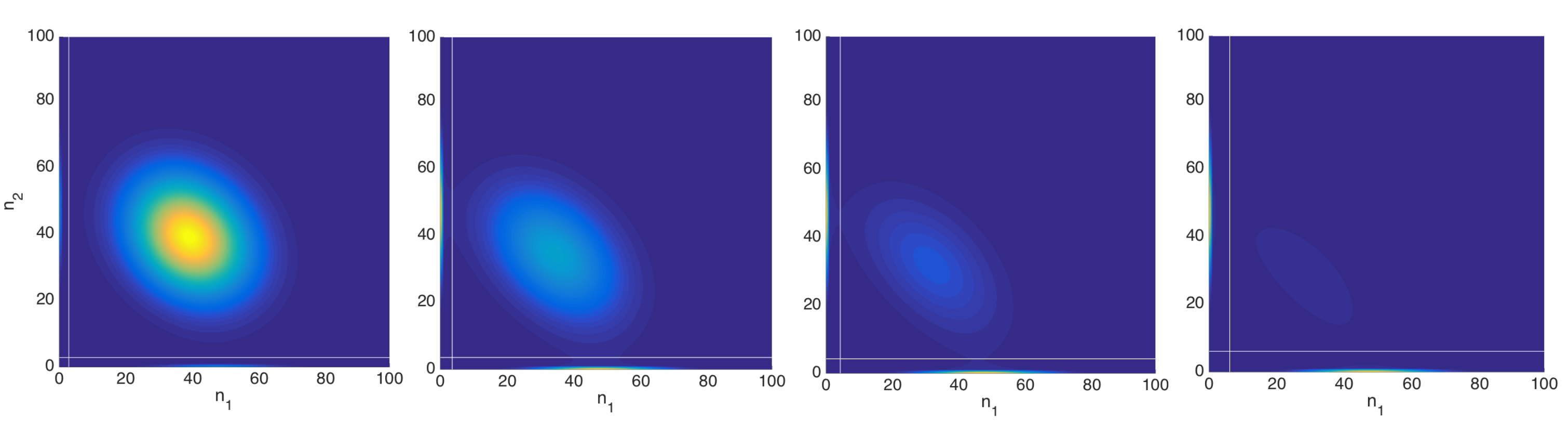}
\caption{
\label{fig:partS2} Partitioning of the configuration space $\mathbb{N}^2$ for $S=2$ 
species for (from left to right) $\rho=0.2$, $\rho=0.35$, $\rho=0.45$ and
$\rho=0.6$; remaining parameters are $r^+=50$, $r^-=27$, $\mu=1$, and $K=50$.
The smaller coordinates of the two saddle points (used to draw white lines) 
are used to partition the space into
regions associated to coexistence (central square), one-species extinction (the
two rectangles which contain the boundary maxima) and two-species extinction (the
small square that contains the origin). As competition increases, the coexistence
maximum approaches to the origin, and saddle points become closer to the
maximum.
}
\end{center}
\end{figure*}

The coordinates for the interior critical points are: on the one hand, $M_1=(m_+,m_+,m_+)$ 
and $M_2=(m_-,m_-,m_-)$, where
\begin{equation}\label{eq:intmaxS3}
m_{\pm}=\frac{K}{2(1+2\rho)}\left(1\pm\sqrt{1-\frac{4\alpha(1+2\rho)}{K}}\right).
\end{equation}
Both solutions turn out to be complex when
\begin{equation}\label{eq:thM}
\rho > \frac{1}{2}\left(\frac{K}{4\alpha}-1\right).
\end{equation}
On the other hand, six interior critical points appear at points $Q_i$, $i=1,\dots,6$, 
where $Q_1=(t_+,t_+,s_-)$ and $Q_2$, $Q_3$ are obtained as the cyclic permutations of the
entries of $Q_1$, whereas $Q_4=(t_-,t_-,s_+)$ and the entries of $Q_5$ and $Q_6$ are the cyclic 
permutations of that of $Q_4$, with
\begin{equation}\label{eq:saddleS3}
\begin{aligned}
&t_{\pm}=\frac{K}{2(1+\rho)}\left(1\pm
\sqrt{1-\frac{4\alpha(1+\rho)}{K(1-\rho)}}\right),\\
&s_{\pm}=\frac{K}{2}\left(1\pm\sqrt{1-\frac{4\alpha(1+\rho)}{K(1-\rho)}}\right).
\end{aligned}
\end{equation}
Both $s_{\pm}$ and $t_{\pm}$ are real whenever $\rho\le\rho_{\mathrm{c}}$, 
where
\begin{equation}\label{eq:th3}
\rho_{\mathrm{c}}=\frac{1-4(r^+-\mu)/Kr}{1+4(r^+-\mu)/Kr}.
\end{equation}
In spite that Eqs.~\eqref{eq:intmaxS3} and \eqref{eq:saddleS3} have been
obtained using an approximate form for conditional probabilities, the
numerical calculation of the critical points using the actual distribution
is in very good agreement with these approximations (see Fig.~\ref{fig:cp}b). 
For $\rho<\rho_{\mathrm{c}}$, $M_1$ is classified as a local maximum. At
$\rho=\rho_{\mathrm{c}}$ all saddle points and $M_1$ coalesce 
in a single point. For $\rho>\rho_{\mathrm{c}}$, however, $M_1$ becomes a 
saddle point, as can be checked numerically with the Hessian matrix.

Boundary maxima are of the form $(n,n,0)$, $(n,0,0)$ ---and their cyclic
permutations. The non-zero coordinates of the former are equal to that of 
the coexistence maximum $M_1$ obtained for $2$ species, see Eq.~\eqref{eq:intmaxS2}; 
the non-zero entries of the latter are the same as the boundary maxima for $S=2$, see 
Eq.~\eqref{eq:bdmaxS2}.

\subsubsection{Configuration-state partitioning}

For $S=2$ potential species, a simple way to divide the configuration space is
to use saddle points. Fig.~\ref{fig:partS2} depicts steady-state distributions for
increasing competitive overlap as well as the location of saddle points. A natural
partitioning should relate configurations around the coexistence maximum to species
coexistence, and states near the boundary maxima to configurations close to one-species 
extinction. As Fig.~\ref{fig:partS2} shows, saddle points discriminate with accuracy
the configurations that can be associated to coexistence from those that can be
related to one-species extinction. According to the coordinate ($s_-$) that
closest to the boundary (cf. Eq.~\eqref{eq:saddleS2}), the partitioning results as:
\begin{enumerate}
\item $0\le n_i<s_-$ for $i=1,2$. This square is associated with full extinction.
\item $0\le n_1<s_-,n_2>s_-$  or $0\le n_2<s_-,n_1>s_-$. These two
rectangles are associated with the extinction of one species, since configurations
are close to extinction in the form $(n,0)$ or $(0,n)$.
\item $n_i>s_-$ for $i=1,2$. The rest of the configuration space puts together states
that can be associated to coexistence.
\end{enumerate}
Note that, strictly speaking, there will be configurations where both $n_1,n_2>0$
are classified as one- or two-species extinction states. The classification here is
meant to separate configurations that are close to boundary maxima in which
one or two species are extinct from those that can be associated to coexistence,
in which species populations are far from being extinct.

\begin{figure}[t!]
\begin{center}
\includegraphics[width=0.9\columnwidth]{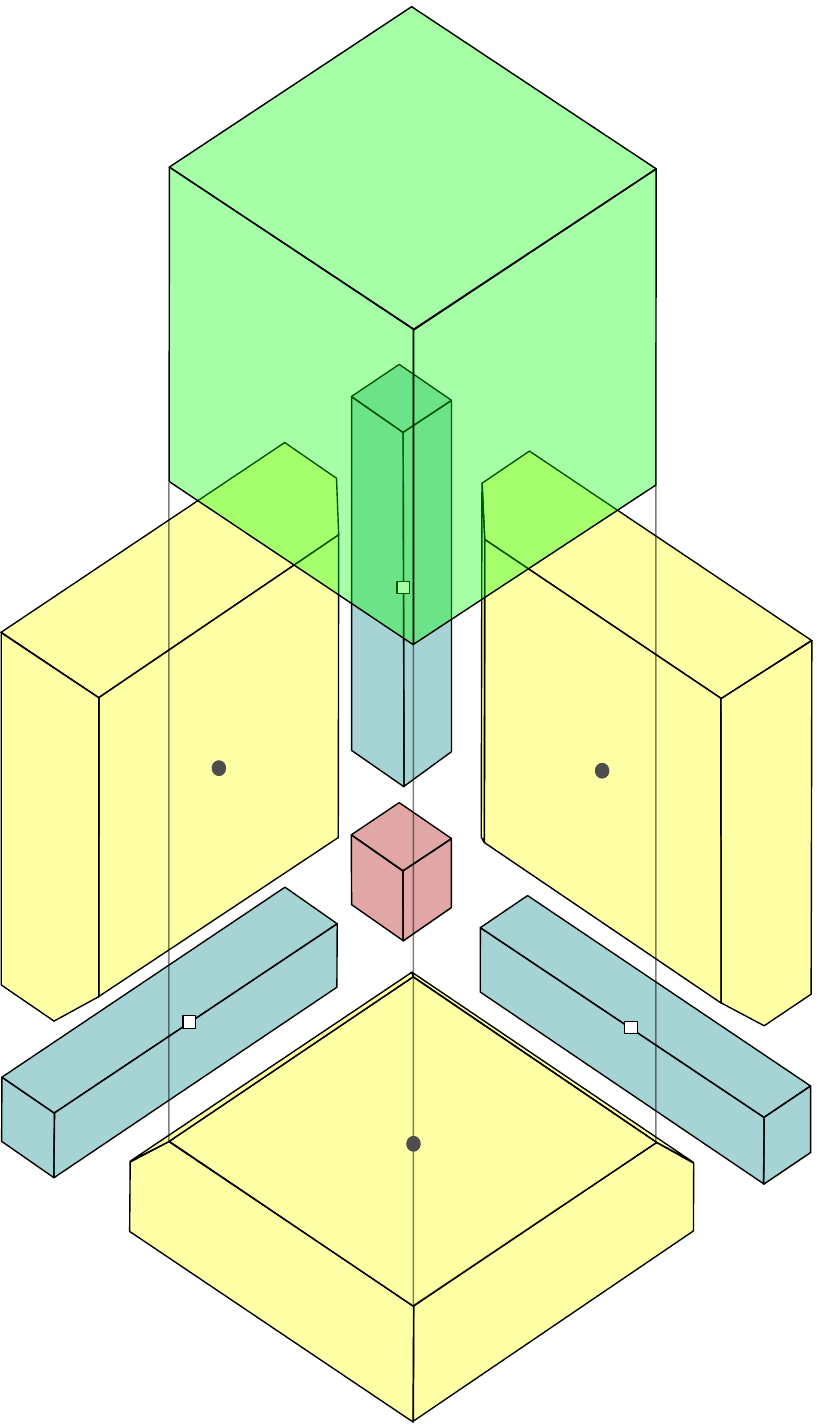}
\caption{\label{fig:part}
Partition of the configuration space $\mathbb{N}^3$ for $S=3$ potential species. To 
ease visualization, we have separated $\mathbb{N}^3$ into four regions according to 
saddle points: black circles denote points $Q_1$, $Q_2$ and $Q_3$, whereas white squares 
represent saddle points $Q_4$, $Q_5$ and $Q_6$. The latter are used to define regions for
complete (red) and two-species extinctions (blue), and the former determine 
the coexistence volume (green) and the one-species extinction region (yellow). 
The (green) cube has been displaced vertically to facilitate visualization.
}
\end{center}
\end{figure}

The partitioning for three-species communities simply generalizes the $S=2$ case. 
Again, each saddle point has a coordinate close to the boundary (cf. $s_-$ and $t_-$
in Eq.~\eqref{eq:saddleS3}). Note also that $s_->t_-$. Taking this fact into
account, saddle points divide the configuration space $\mathbb{N}^3$ into
several regions, which have been depicted in Fig.~\ref{fig:part}:
\begin{enumerate}
\item $0\le n_i<t_-$ for $i=1,2,3$. This cube is associated with the extinction of the
three species (since the origin belongs to this region).
\item $0\le n_i<t_-$ for $i=1,2$ and $n_{3}>t_-$ (and the two remaining combinations). 
These three parallelepipeds are associated with the extinction of two species, because
boundary maxima of the form $(n,0,0)$ ---and its cyclic permutations--- are situated 
inside this volume, as well as other population configurations close to two-species 
extinctions.
\item $n_i>s_-$ for $i=1,2,3$. This cube contains the interior maximum and is 
therefore associated with the coexistence of the three species.
\item The remaining volume of the configuration space, where boundary maxima
of the form $(n,n,0)$ ---and its cyclic permutations--- are located, is associated with 
the extinction of one species.
\end{enumerate}

The configuration-state partitioning slightly differs from the general case
when saddle points have coalesced. If $\rho>\rho_{\mathrm{c}}$,
the point $M_1=(m_+,m_+,m_+)$ is classified as the only saddle point. 
Based on its coordinates, the configuration space is partitioned as follows:
\begin{enumerate}
\item $0\le n_i<m_+$ for $i=1,2,3$. This cube is associated with full extinction.
\item $0\le n_i<m_+$ for $i=1,2$ and $n_{3}>m_+$ (and the two remaining combinations). These
three parallelepipeds are associated with the extinction of two species.
\item $n_i>m_+$ for $i=1,2,3$. This cube is associated with coexistence configurations.
\item The remaining volume of the configuration space is related to the extinction of 
one species.
\end{enumerate}
The same partitioning applies for the $S=2$ case when only a single saddle point
remains.

\subsection{Extinction cascade}
\label{sec:cascade}

\begin{figure*}[t!]
\begin{center}
\includegraphics[height=6.8cm]{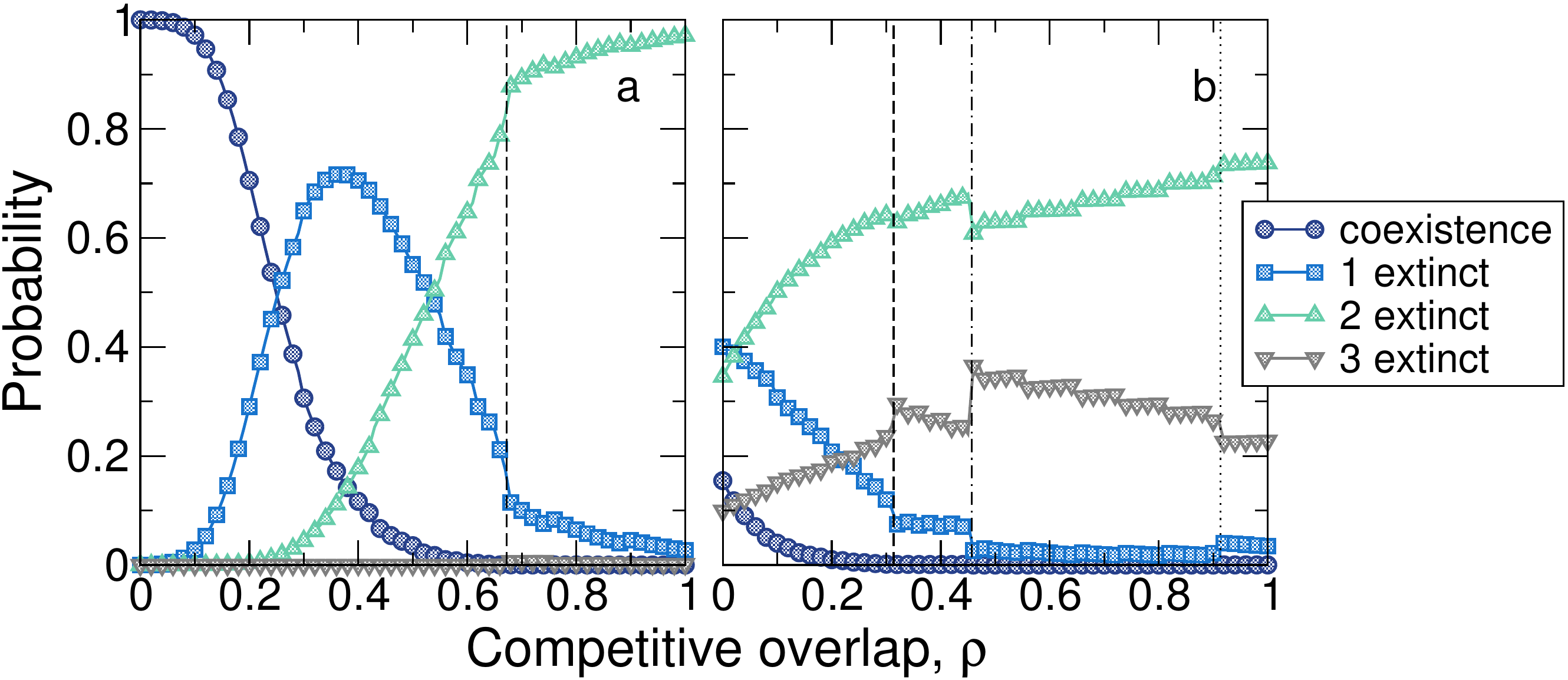}
\caption{
\label{fig:coexprob} 
(a) Stochastic extinction cascade. The panel shows the dependence of
aggregated probabilities over the regions determined by the critical points
as a function of competitive overlap $\rho$. Parameter values are 
$S=3$, $r^+=50$, $r^-=10$, $K=25$, and $\mu=1$. Although initially the three
species coexist, at intermediate values of $\rho$ the most probable state is 
formed by only two extant species. For large competition values the most likely
configuration comprises a single extant species. A vertical, dashed line marks 
the value $\rho_{\mathrm{c}}$ (cf. Eq.~\eqref{eq:th3}) at which the probability 
of coexistence is negligible. (b) Same as panel a but for a higher intrinsic mortality 
rate, $r^-=35$. In this case, the aggregated probability of complete extinction is 
non-negligible even for $\rho=0$. Aggregated probability for one-extinct species 
configurations starts declining and, at the same time, the two-extinct species 
configurations become more likely as $\rho$ increases. Close to $\rho=1$ the 
system alternates most of the time between a single-species state or a completely 
extinct community. The vertical, dashed line marks the threshold $\rho_{\mathrm{c}}$.
The dot-dashed line shows the value given by Eq.~\eqref{eq:thM}, at which
the two critical points $M_1$ and $M_2$ no longer exist. Finally, a dotted, vertical line
marks the value ($\rho=K/4\alpha-1$, see Section~\ref{sec:cp2}) at
which the boundary maxima of the form $(n,n,0)$ ---and permutations---
no longer exist.
}
\end{center}
\end{figure*}

Saddle points of the joint probability distribution have allowed us to establish a 
natural partitioning into regions associated to coexistence (containing
the coexistence maximum) and to the extinction on one, two, or three
species (containing the corresponding boundary maxima), see Fig.~\ref{fig:part}. Over 
these regions we aggregate the joint distribution $P(n_1,n_2,n_3)$, calculated 
numerically as described in~\ref{sec:appB}, to define the overall 
probability of coexistence,
\begin{equation}
P_{\mathrm{c}}=\sum_{n_1,n_2,n_3>s_-}P(n_1,n_2,n_3),
\end{equation}
the probability of three-extinct species configurations,
\begin{equation}
P_3=\sum_{n_1,n_2,n_3<t_-}P(n_1,n_2,n_3),
\end{equation}
and the probability of two-extinct species, which by symmetry over cyclic
permutations of $(n_1,n_2,n_3)$ can be expressed as
\begin{equation}
P_2=3\sum_{n_1,n_2<t_-}\sum_{n_3>t_-}P(n_1,n_2,n_3).
\end{equation}
The probability of one-extinct species, $P_1$, is obtained from the
normalization condition $P_{\mathrm{c}}+P_1+P_2+P_3=1$. 
Fig.~\ref{fig:coexprob} shows these aggregated probabilities as a function of
$\rho$ for two sets of model parameters. In the first case, the coexistence
probability is almost one for low values of the competitive overlap and, as
$\rho$ increases, at some point the probability declines rapidly. At the
same time, the probabilities of one and two extinct species begin to 
increase. Note that, once the threshold $\rho_{\mathrm{c}}$ has been 
crossed over, the most probable state consists of a single, extant species,
and the probability of coexistence becomes negligible.

In the second case, corresponding to a larger value of the mortality rate 
$r^-$, the threshold $\rho_{\mathrm{c}}$ at which the coexistence 
maximum $M_1$ transforms into a saddle point becomes smaller. The 
probability of coexistence rapidly declines as $\rho$ increases and, in 
addition, there is a non-negligible probability of complete extinction.
Remarkably, Fig.~\ref{fig:coexprob}b shows that, for smaller values of 
the carrying capacity, coexistence is not the only possible state even at 
$\rho=0$. This puts a practical limit to the maximum number of coexisting 
species which does have a deterministic counterpart ---recall that, due
to global stability, the deterministic model permits the packing of an 
arbitrary number of species for $\rho<1$.

Contrary to the deterministic prediction that $S-1$ extinctions take place 
abruptly as $\rho$ increases (\ref{sec:appA}), Fig.~\ref{fig:coexprob} shows
that ecological drift induces a sequential cascade of extinctions, in which
states with a larger number of extinct species are more prone to be 
observed as the competitive overlap increases.

An important remark is on purpose here. The cascade of extinction we have
just described has nothing to do with the degree of synchronicity in which 
extinctions take place along time, i.e., the term ``cascade'' does not refer 
here to a sequential extinction in time. In particular, the symmetric, deterministic 
model leads in general to asynchronous extinctions. The stochastic 
phenomenon analyzed in this contribution refers to the progressive extinctions 
that occur as competitive strength increases.

\subsection{Limit of small demographic stochasticity}
\label{sec:demnoise}

In the absence of stochasticity, the deterministic model predicts the extinction of $S-1$
species once the threshold in competition $\rho=1$ is crossed over. In the stochastic
case, the extinction threshold is pushed to smaller values of competitive 
strength~\citep{capitan:2015}, and
the probabilities of configurations with one or more extinct species are non-zero in
overlapping windows of competition. These two scenarios only differ on the presence or 
absence of demographic stochasticity, but lead to significantly different outcomes. 
Therefore, incorporating demographic stochasticity appears to be very relevant in 
the dynamics of ecological community models. 

In order to evaluate the importance of demographic stochasticity, we have tried to
quantify the difference between these two scenarios as stochasticity decreases. To do
so, we have studied the limit of small stochasticity, in which the deterministic model is 
to be recovered. As shown below, the transition to the small-stochasticity scenario is abrupt, 
hence incorporating demographic stochasticity to community models should be strongly 
considered. 

Since fluctuations are expected to decrease as population size increases (see~\ref{sec:appC}),
the small-stochasticity limit is equivalent to the limit of large population sizes, so we
have repeated the analysis by increasing the carrying capacity at fixed $\rho$. 
We have quantified the intensity of demographic noise by the 
coefficient of variation of population abundances, $\nu=\sigma_n/\langle n\rangle$, $\sigma_n$
being the standard deviation of population numbers and $\langle n\rangle$ the average value. As
shown in~\ref{sec:appC}, when $K\gg1$ then $\sigma_n\sim K^{1/2}$ and $\langle n\rangle\sim K$, 
so $\nu$ tends to zero in the limit of large population sizes. From the numerical point of view, 
to get close to the deterministic scenario we would have to choose a 
carrying capacity value such that the average $\langle n\rangle$ is large enough compared
to the variability in populations. In practice, we will assume that the system is
close to a low-noise regime when the actual coefficient of variation, obtained though
the joint probability distribution of the stochastic model calculated numerically, is close to
that obtained by a Gaussian approximation of the joint distribution valid in the limit $K\gg 1$
(see~\ref{sec:appC}). Note that, for the Gaussian approximation to be valid, the coexistence,
interior maxima must be located far away from the boundaries, so that the joint probabilities 
associated to all of extinction states are negligible.

\begin{figure}[t!]
\begin{center}
\includegraphics[width=\columnwidth]{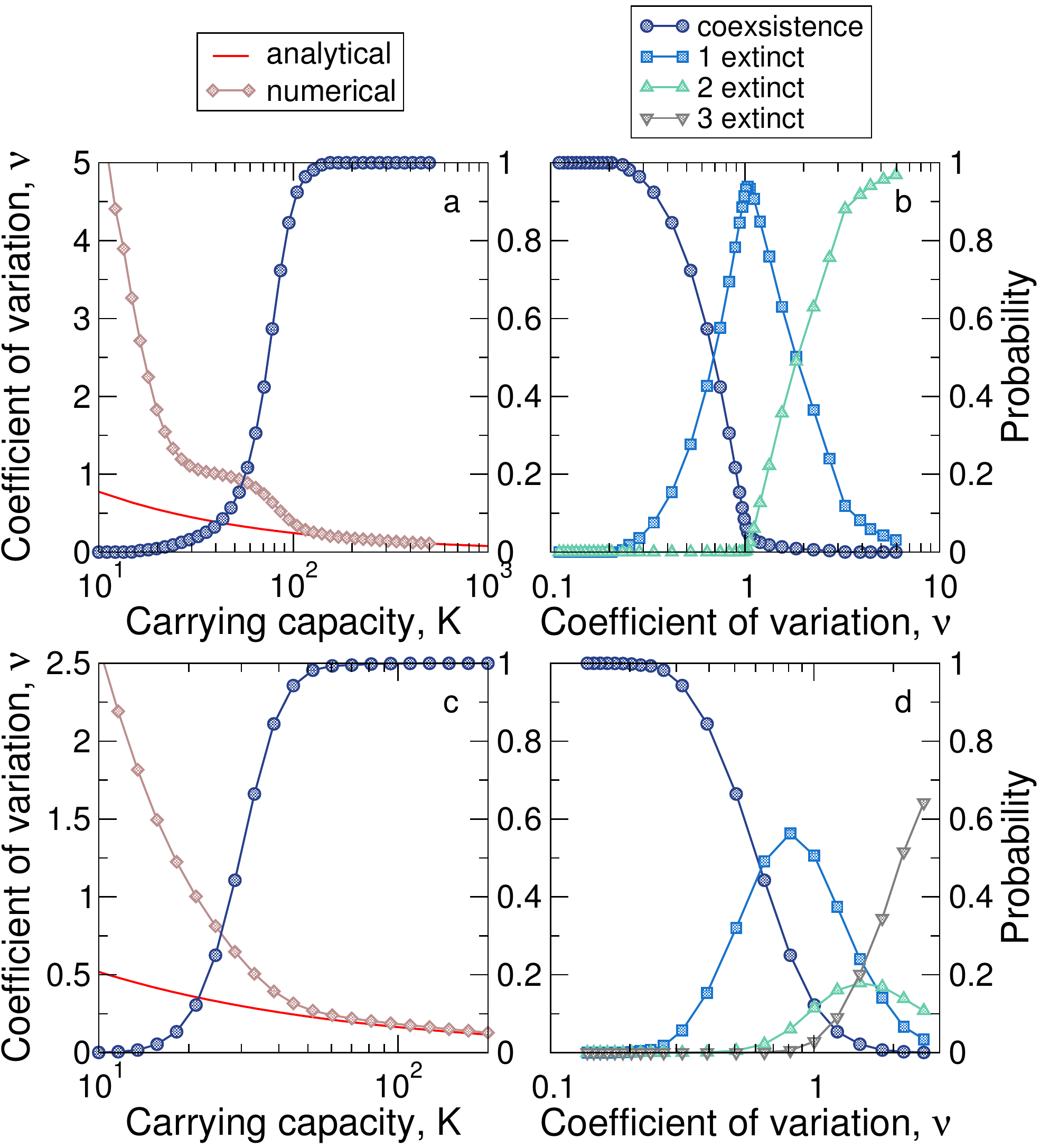}
\caption{
\label{fig:demnoise} Cascade persistence for variable levels of demographic 
stochasticity. Panel (a) shows the coefficient of variation of population abundances, 
$\nu=\sigma_n/\langle n\rangle$ (diamonds), as a function of increasing carrying capacity, 
$K$, for $S=2$, $r^+=50$, $r^-=25$, $\mu=0.1$ and $\rho=0.5$. For large $K$, 
the coefficient of variation tends to the analytical approximation~\eqref{eq:CV} 
derived in~\ref{sec:appC}. Simultaneously, the probability of coexistence 
(alternative vertical axis, circles) becomes closer to $1$. (b) Extinction cascade
as a function of demographic stochasticity. At $\nu\sim 0.2$, the probability that
one species goes extinct abruptly increases, and the coexistence probability
starts declining. As $\nu$ increases, higher-order extinctions become more likely. 
Panels (c)-(d) are equivalent to (a)-(b) but for $S=3$, $r^+=10$, $r^-=5$, $\mu=0.1$
and $\rho=0.1$. When the Gaussian approximation is valid, coexistence probability
is almost equal to $1$. After an abrupt decline, sequential extinctions occur for higher 
levels of demographic noise.
}
\end{center}
\end{figure}

The results of this analysis are summarized in Fig.~\ref{fig:demnoise}. For low levels of stochasticity,
the numerical coefficient of variation and the analytical approximation, Eq.~\eqref{eq:CV}, remain 
close to each other. For the corresponding values of $K$, the probability of coexistence is almost
equal to one (as in the deterministic scenario). However, as $\nu$ increases (i.e., the carrying 
capacity $K$ decreases), the probability that one species becomes extinct grows sharply
and, at the same time, the probability of coexistence starts declining. As the
variability in population sizes augments, overlapping windows for the likelihood of progressive
extinctions arise. The transition to situations where low demographic stochasticity operates is,
therefore, abrupt.

Note that the cascade obtained in Fig.~\ref{fig:demnoise}b,d as a function of noise can be 
immediately translated into a cascade in carrying capacity (for fixed $\rho$). This reinforces
our conclusion, since the extinction cascade phenomenon also occurs when other model 
parameter ($K$) varies. It is presumably the relative balance between $\rho$ and $K$ that
determines the subspace of the parameter space for which extinctions start appearing. 

\subsection{Evaluating the role of environmental stochasticity}
\label{sec:envnoise}

In order to test the robustness of our main result against different sources of noise, 
we have replaced demographic stochasticity
by environmental stochasticity. We have introduced variability in model parameters so that,
to keep the scenario as simple as possible, the competitive overlap $\rho$ in Eq.~\eqref{eq:dyn}
is replaced by $\rho+\xi(t)$, where $\xi(t)$ stands for a noise term with zero mean. The deterministic
dynamics transforms into a Langevin equation with multiplicative noise,
\begin{equation}\label{eq:langevin}
\dot{x}_i=rx_i\left(1-\frac{x_i+\rho\sum_{j\ne i}x_j}{K}\right)-\frac{r x_i}{K}\bigg(\sum_{j\ne i}x_j\bigg)\xi(t),
\end{equation}
which can be rewritten as
\begin{equation}\label{eq:scaledlang}
z_i'=z_i\bigg(1-z_i-\rho\sum_{j\ne i}z_j\bigg)-z_i\bigg(\sum_{j\ne i}z_j\bigg)\xi(t)
\end{equation}
by re-scaling species densities as $z_i=x_i/K$ and time as $t'=rt$ ($z_i'$ stands for the derivative with
respect to the scaled time $t'$). We choose the noise as $\xi(t)=\rho\kappa\eta(t)$, where $\eta(t)$ is
a Brownian motion. The noise $\xi$ has been scaled by $\rho$ in order to avoid that the overall
competitive strength, $\rho+\xi(t)$, becomes negative.

\begin{figure*}[t!]
\begin{center}
\includegraphics[height=6.8cm]{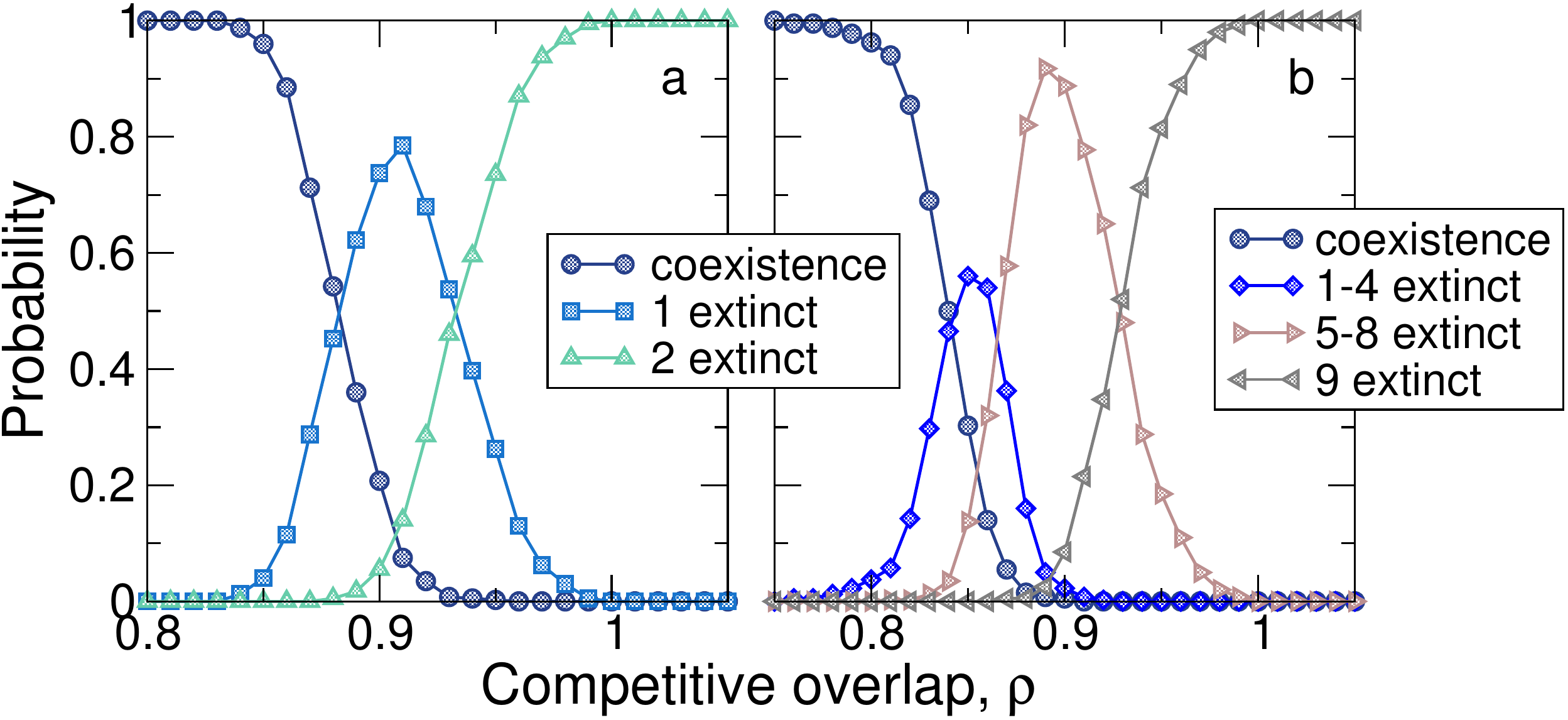}
\caption{
\label{fig:envnoise} Cascade of extinctions in the case of environmental stochasticity. The Langevin
equation~\eqref{eq:scaledlang} has been numerically integrated for (a) $S=3$ and (b) $S=10$ species.
At the end of the simulated time span, species $i$ is regarded as extinct if the corresponding density 
$z_i$ is at least $1\%$ smaller than the maximum density $z_{\mathrm{m}}=\max\{z_1,\dots,z_S\}$. 
Probabilities are calculated here by averaging over $400$ realizations starting from random initial 
conditions. Here we take $\kappa=0.5$
in the definition of the multiplicative noise $\xi$. To ease visualization, in panel (b) we have aggregated
together the probabilities from one to four extinctions, as well as the probabilities of observing from
five to eight extinctions. Consistently with the deterministic model, the full extinction state is never 
observed even in the presence of noise.
}
\end{center}
\end{figure*}

Fig.~\ref{fig:envnoise} shows the persistence of the extinction cascade when only environmental
stochasticity is considered. This points to the robustness of our results: as for demographic stochasticity,
environmental stochasticity also alters the predictions of the deterministic dynamics. There are, however,
qualitative differences between the predictions yielded by the model when demographic or environmental 
stochasticity come into play. First, the range in competition on which the cascade takes place is wider for 
demographic noise. It seems that, in the presence of environmental noise, the range of the cascade can 
increase moderately when a larger number of species are to be packed (Fig.~\ref{fig:envnoise}b). 
More importantly, a second difference arises: no full extinction is possible in the case of environmental
noise. This is a peculiarity, not altered by the noise, of a generic competitive, Lotka-Volterra dynamics
(cf. Eq.~\eqref{eq:detdyn} in~\ref{sec:CEP}), for which it can be easily shown that the full extinction 
equilibrium ($\hat{x}_i=0$, $i=1,\dots,S$) is unstable. The Langevin equation, therefore, can not reproduce 
configurations associated to the full extinction of the community, contrary to what is observed for 
demographic stochasticity (Fig.~\ref{fig:coexprob}b).

\subsection{Relaxing the species symmetry assumption}
\label{sec:nosym}

In this section we test the robustness of our results by relaxing the assumption of species symmetry
in model parameters. It can be argued that the effect of demographic stochasticity simply consists of 
breaking the symmetry between species. In a generic, non-symmetric, deterministic
scenario, one could expect progressive extinctions even in the absence of ecological drift. Therefore,
the role of stochasticity would be simply to re-establish a deterministic scenario where one-by-one
species extinctions occur. In this subsection we discuss the implications of relaxing the symmetry 
assumption to determine the true role of demographic stochasticity in a generic case.

In order to address these questions, here we consider two examples of fully non-symmetric, three-species 
competitive dynamics,
\begin{equation}\label{eq:nosym}
\dot{x}_i=rx_i\left(1-\frac{1}{K_i}\sum_{j=1}^3\rho_{ij}x_j\right),
\end{equation}
where carrying capacities and interspecific competitive strengths are species-dependent. We set, 
without loss of generality, off-diagonal competition values as $\rho_{12}=\rho_{21}=\rho$, 
$\rho_{13}=\rho_{31}=\rho+\delta_1$, $\rho_{23}=\rho_{32}=\rho+\delta_2$, and $\rho_{ii}=1$ 
for $i=1,2,3$. 

\begin{figure*}[t!]
\begin{center}
\includegraphics[height=6.8cm]{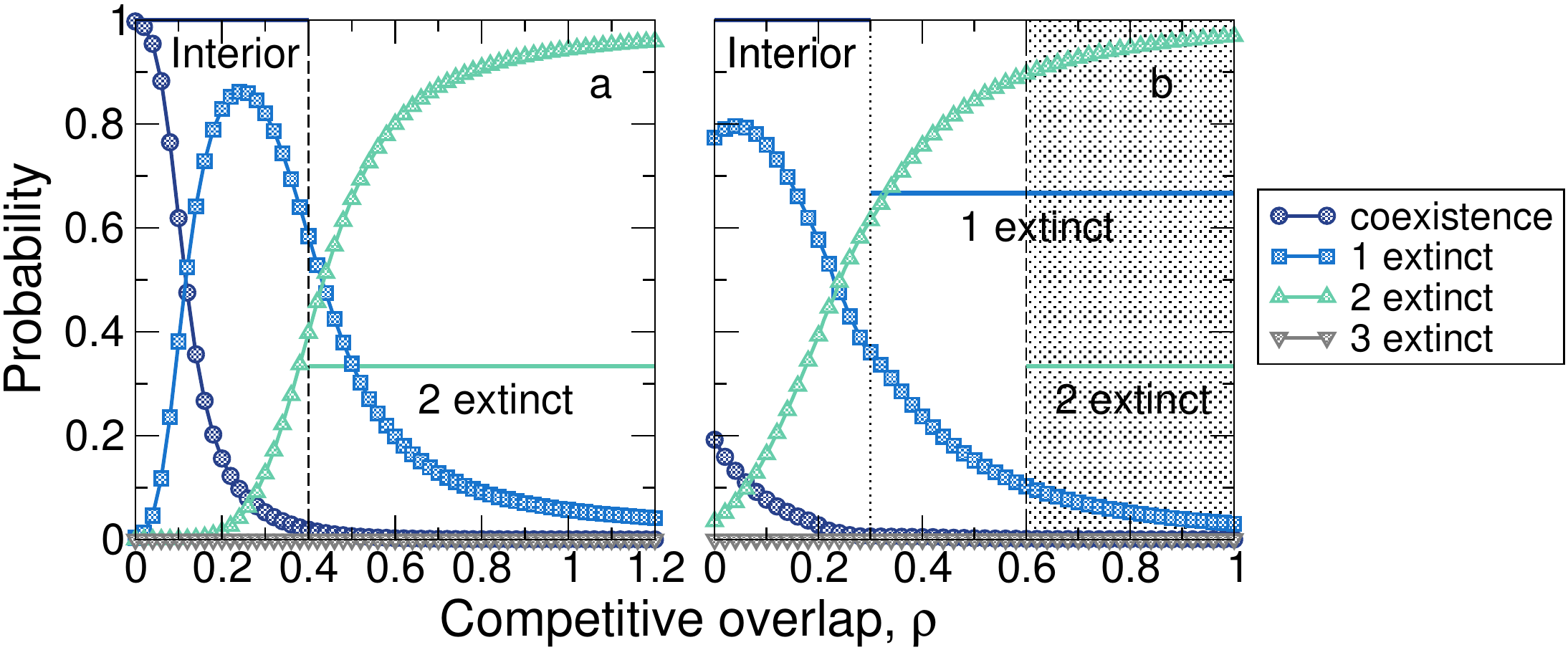}
\caption{
\label{fig:nosym} Non-symmetric scenario described in Sec.~\ref{sec:nosym}. In both panels,
$r^+=5.1$, $r^-=0.1$ and $\mu=0.1$. (a) In this case, the deterministic model predicts a multiple
extinction of two species when the threshold $\rho=0.4$ is crossed over (the ranges where the 
deterministic model yields stability are marked with horizontal lines). Remaining parameters are 
$K_1=40$, $K_2=16$, $K_3=20$, $\delta_1=0.1$ and $\delta_2=0.05$. The stochastic scenario, 
however, yields a sequential cascade for which extinctions are spread out along the axis of 
competitive overlap (symbols). (b) Here carrying capacities are uniform
($K_1=K_2=K_3=15$) and $\delta_1=0.4$ and $\delta_2=0.3$. The stability analysis of the 
deterministic model implies ranges where multiple stable equilibria co-occur (shadowed area).
The extinction sequence is not fully determined, and has to be compared with the stochastic 
prediction.
}
\end{center}
\end{figure*}

In the first example we choose $\delta_1=0.1$, $\delta_2=0.05$, $K_1=40$, $K_2=16$, and $K_3=20$.
After analyzing the stability of all the equilibrium points of the deterministic system (see details 
in~\ref{sec:appD}) we find that, although being fully non-symmetric, this model predicts a two-species,
grouped extinction when the threshold $\rho=0.4$ is crossed over (Fig.~\ref{fig:nosym}a). For 
$\rho>0.4$, only equilibria with a single extant species are stable, whereas no other scenario but three-species
coexistence is stable for $0\le\rho<0.4$. By continuity of the eigenvalues of the Jacobian matrix on
model parameters, there are multiple non-symmetric, deterministic scenarios that exhibit the transition 
from full coexistence to a single-extant-species state. Therefore, grouped extinctions are not a peculiarity
of the fully symmetric scenario.

In the second example (Fig.~\ref{fig:nosym}b), $\delta_1=0.4$, $\delta_2=0.3$ and $K_1=K_2=K_3=15$.
The complete stability analysis draws the conclusion that non-symmetric models yield to 
multistability. In~\ref{sec:appD} it is shown that, for $\rho<0.3$, only the interior equilibrium point is 
asymptotically stable; for $0.3<\rho<0.7$, only the boundary point $(15/(1+\rho),15/(1+\rho),0)$ is
stable, but for $0.7<\rho<1$, the former boundary point remains stable as well as the single-species
extinction equilibrium $(0,0,15)$ (see also Fig.~\ref{fig:nosym}b). For $\rho>1$, however, 
the three boundary equilibria with a single, extant species are the only ones that remain stable. 
As a consequence, there is a range in competition ($0.7<\rho<1$) where configurations formed by a 
single extant species or by two coexisting species co-occur. Depending on the initial
condition, the dynamics can end up in one of the two attractors. The basin
of attraction of each equilibrium point will determine how frequently one or two species go extinct
when competition surpasses the value $\rho=0.7$. This analysis is out of the scope of this contribution, 
though. Importantly, the extinction sequence in non-symmetric scenarios can depend on initial 
conditions and is not fully determined in principle.

Multiple extinctions are not precluded in general even when the symmetry between species is broken,
as the first example shows. Multistability ranges could also lead to grouped extinctions in deterministic
scenarios. In a Lotka-Volterra community model with $S$ species there are $2^S$ attractors. 
Increasing complexity would likely lead to additional overlapping regions in competition where 
multiple stable attractors co-occur and species can decline together. Moreover, the deterministic
extinction can be ambiguous. It seems difficult to establish the conditions under which a general 
non-symmetric model will produce a well-defined extinction sequence. The variability introduced by 
idiosyncratic, species-dependent carrying capacities, growth rates, or intra- and interspecific strengths, 
may cause the extinction sequence to be analytically unpredictable for species-rich communities.

Remarkably, the extinction sequence predicted by non-symmetric, deterministic models 
has nothing to do with the cascade observed when demographic stochasticity comes into
play. We have calculated the probabilities of coexistence and one-, two- or three-species extinctions
for both stochastic parametrizations of the two non-symmetric models considered in this subsection.
To aggregate joint probabilities, we obtained numerically the critical points by
spline interpolation of the exact joint distribution. Now the six saddle points have asymmetric entries,
but three of them exhibit a coordinate close to the boundary, and the remaining three saddle points
present two coordinates near the axes. Thus, the partition of the configuration space is conceptually 
equivalent to that of Fig.~\ref{fig:part}. Not surprisingly, as Fig.~\ref{fig:nosym} evidences, the 
stochastic model predicts a sequential cascade of extinctions. The threshold at which 
extinctions start occurring displaces towards smaller values of $\rho$, leading to narrow ranges 
of effective stochastic coexistence. The most likely state in the
presence of stochasticity is not necessarily the same as in a deterministic scenario, the predictions 
being utterly different in terms of the extinction sequence. Therefore, regardless of the inherent lack 
of symmetries that deterministic dynamics may have, demographic stochasticity can influence 
significantly the way in which extinctions take place.

\section{Discussion}
\label{sec:discussion}

In this contribution, we have analyzed the extinction phenomenon for 
a symmetric, Lotka-Volterra competitive system formed by $S$ species. In 
particular, we have focused on the differences between the deterministic system
and its stochastic counterpart. Our main result is related to the way in which
species extinction proceeds: on the one hand, in the deterministic system, $S-1$ 
species are driven to extinction at the very point where competitive exclusion starts 
to operate. On the other hand, we have shown that the overall probabilities of 
coexistence and one-, two-, or three-species extinction alternate
sequentially as the most likely states when competitive overlap increases.
Therefore, stochasticity is responsible of a progressive sequence of species
extinction, a phenomenon that is absent in the deterministic system. Our analyses
are based on a birth-death-immigration stochastic dynamics that was 
analyzed in deep by~\cite{haegeman:2011} and later used 
by~\cite{capitan:2015} to unveil the existence of a more restrictive threshold 
in competition when demographic stochasticity is explicitly considered.
In addition to lowering the threshold in competition at which extinctions
begin to occur, ecological drift also changes drastically 
the way those extinctions take place. In order to evidence this difference, we 
have developed convenient analytical approximations to the critical points
of the joint probability for $S=2$ and $S=3$ potential species, which
were used to divide the configuration space of the stochastic process to yield
aggregated probabilities associated to coexistence and to the corresponding
configurations with one or more extinct species. These probabilities reveal
the stochastic extinction cascade.

We have shown that different community configurations alternate within 
certain ranges of competition: coexistence and one-species extinction alternate 
for small $\rho$, whereas one-species and two-species extinction most 
likely interchange among each other for intermediate values of $\rho$. 
The presence of multiple modes in the joint probability distribution is akin 
to the presence of multistability in a deterministic system. However, 
the symmetric, deterministic model is characterized by the absence of multiple 
stable states for $\rho<1$. Hence the extinction cascade described in this 
work is an entirely new effect caused only by ecological drift. In addition, 
we have evidenced that the transition to the deterministic model is sharp when
the intensity of demographic stochasticity tends to zero. Moreover, we have
proven that environmental stochasticity also leads to a cascade of extinctions,
although the ranges in competition where extinctions take place are smaller and,
more importantly, the full extinction of the community is not possible in this case.

In this work, we have implemented two types of stochasticity: demographic 
noise (ecological drift) and environmental stochasticity. These are two typical 
sources of noise that represent, respectively: (i) the variability in discrete 
population numbers as a consequence of stochastic births and deaths, or
(ii) the stochastic variability in model parameters that can be ascribed 
to changing environmental conditions. Although they are very different 
implementations of noise, the main result of this manuscript (the stochastic 
cascade of extinctions) is common to both of 
them, with some qualitative differences. In the case
of demographic stochasticity, we do not impose a particular form for the
noise distribution since it directly emerges from the inherent stochastic dynamics of 
discrete populations whose individuals undergo a number of 
elementary processes (in principle, the noise distribution would follow from the 
master equation). Other ways to implement demographic noise have been
discussed in the literature~\citep{bonachela:2012}, and they would plausibly lead to
mechanisms similar to those found here.

It can be argued that the role of stochasticity in the fully symmetric system 
reduces to break species symmetry and yield to progressive species extinctions,
a scenario that can arise in non-symmetric, deterministic approaches.
We have illustrated with examples that grouped extinctions are not exclusive
of a fully symmetric situation. Moreover, predicting the extinction sequence
in non-symmetric, deterministic cases is difficult because multiple stable equilibria can
co-occur in ranges of competition. Besides, although the extinction sequence
were completely determined, the cascade in the presence of stochasticity can
be totally different from that predicted by the deterministic model. We believe
that the examples analyzed in this contribution show up the key role of
stochasticity in community assembly.

It would be interesting to empirically test the stochastic extinction cascade
phenomenon. In principle, a plausible way to conduct the experiment would
involve simple protist microcosms where species compete for a shared resource 
(see, for example,~\cite{violle:2010} and references therein). Lowering the amount
of resource could be associated to a decrease in the carrying capacity, and we have
shown that the extinction cascade mechanism is expected to arise as long as the 
carrying capacity (resource availability) is reduced, see Fig.~\ref{fig:demnoise}. 
For the experiment to reproduce 
model settings, an individual (immigrant) coming from the species pool should be
inserted in the experimental community at certain times. From a time series listing 
species identities at certain sampling times, one could estimate the probabilities for 
coexistence and for the extinction of one or more species, and test whether extinctions 
in empirical systems tend to proceed sequentially or not.

Our work has two important implications: first, we have developed analytical
approximations for conditional probabilities in the cases $S=2$ and $S=3$.
As we have shown, these functions work well at least around the 
critical points of the joint probability. Presumably, the techniques 
proposed here might be extended to approximate the joint probability
distribution itself. This approach, however, has to be performed carefully.
The simplest way to approximate the three-species joint distribution according
to our methodology is to set $P(n_1,n_2,n_3)\approx T(n_1|n_2,n_3)T(n_2|n_3)P(n_3)$,
where $P(n_3)$ is the marginal, one-species probability distribution, which can
be expressed analytically in terms of hypergeometric functions~\citep{haegeman:2011}. 
Apparently, the approximated distribution lacks of an important property of the exact 
joint distribution: it is not conserved under cyclic permutations of its
arguments. Therefore, it is necessary to devise appropriate combinations
of conditional probabilities that preserve the symmetry of the joint distribution 
under cyclic permutations of its arguments. This research direction, together 
with a generalization for communities formed by more than three species, could 
be worth pursuing and compared with other approximations to the 
joint probability, such as those developed by~\cite{haegeman:2011}.

The second implication of this contribution is that we remark the importance
of explicitly considering ecological drift in theoretical frameworks in community 
ecology. Natural processes are intrinsically stochastic, because changes in 
population numbers are discrete, so ecological communities are more reliably 
modeled using stochastic community models, even at regimes where their 
deterministic limits are not expected to fail. In certain situations, the use of 
deterministic dynamics in community ecology could lead to utterly different predictions, 
as we have shown. Traditionally ecology has relied on these kind of models, 
and currently considerable theoretical progress is made on the basis of 
deterministic approaches. However, ecological drift may play a determinant 
role that could not be captured by deterministic formulations. We hope that the 
present work can inspire new contributions in the future that highlight the distinct 
role of ecological drift in species community models.

\section{Acknowledgements}
We thank the constructive criticisms and comments of the Editor and two 
anonymous reviewers, who helped to improve significantly the original manuscript.
This work was funded by project SITES (CGL2012-39964, DA, JAC) and the 
Ram\'on y Cajal Fellowship program (DA). JAC also acknowledges partial 
financial support from the Department of Applied Mathematics (Technical 
University of Madrid).

\appendix

\section{Deterministic competitive exclusion}
\label{sec:CEP}

Gause's competitive exclusion principle is usually stated as
``two species competing for a single resource cannot coexist''. Strictly
speaking, the competitive exclusion principle was first formulated
by~\cite{volterra:1926} as a mathematical proposition. Albeit a mathematical 
proof of the principle, based on dynamical systems theory, can be found in the 
book by~\cite{hofbauer:1998}, and is essentially the same as that 
of~\cite{volterra:1926} original paper, we here provide a purely algebraic 
alternative demonstration (see~\cite{roughgarden:1979} as well).

Assume that the densities of $S$ species living on $R$ resources vary in time 
according to the dynamics
\begin{equation}\label{eq:HS}
\dot{x}_i = x_i\Bigg(\sum_{j=1}^R \gamma_{ij} y_j-d_i\Bigg),\qquad i=1,\dots,S,
\end{equation}
where $\mathsf{\Gamma}=(\gamma_{ij})$ is a $S\times R$ matrix with non-negative
entries, $y_j$, $j=1,\dots,R$, are amounts of $R$ resources,
which are also assumed to depend on species densities, and $d_i$, $i=1,\dots,S$ 
are the rates of decline of species when all resources are zero. We now show that
if $S>R$ and species densities reach a well-defined steady state, then in the long 
run $S-R$ species will go extinct until the number of species equals the number 
of resources.

Imposing the condition $\dot{x}_i=0$ for large $t$ and assuming that all species 
densities are positive, we get the non-homogeneous linear system
$\mathsf{\Gamma}\vv{y}=\vv{d}$, with $\vv{y}=(y_i)\in\mathbb{R}^R$, and 
$\vv{d}=(d_i)\in\mathbb{R}^S$. Given that matrix $\mathsf{\Gamma}$ has 
$S$ rows and only $R<S$ columns, the system will be incompatible except for 
some specific vectors $\vv{d}$ belonging to the image 
$\text{Im}(\mathsf{\Gamma})$ of matrix $\mathsf{\Gamma}$. 
As a consequence, to find an equilibrium solution some species densities must go to zero. 
For those displaced species, the corresponding rows of matrix $\mathsf{\Gamma}$ 
can be removed until $\text{rank}(\mathsf{\Gamma})$ rows remain. If the rank 
equals the number $R$ of resources, the system turns out to be compatible and 
determinate, and $R$ species will stably coexist. Note that it is the rank of matrix 
$\mathsf{\Gamma}$ rather than the number of resources itself that induces 
competitive exclusion.

Therefore, competition for shared resources imposes a limit to the maximum 
number of species that can stably coexist. This result has a counterpart in the 
stability of Lotka-Volterra equations derived from the MacArthur's 
consumer-resource model~\citep{macarthur:1970,chesson:1990}, i.e., the 
particular case of Eq.~\eqref{eq:HS} in which per-capita resource growth rates 
$\dot{y}_i/y_i$ depend linearly on population densities. In the limit of fast 
resource variation, the dynamics~\eqref{eq:HS} takes the Lotka-Volterra 
form~\citep{chesson:1990},
\begin{equation}\label{eq:detdyn}
\dot{x}_i=r_ix_i\left(1-\frac{x_i+\sum_{j\ne i}\rho_{ij}x_j}{K_i}\right),
\qquad i=1,\dots,S.
\end{equation}
Here $r_i$ is interpreted as the intrinsic growth rate of species $i$, $K_i$ 
as a carrying capacity, and $\rho_{ij}$ measures interspecific competition strength 
between species $i$ and $j$ relative to intraspecific competition. 
As shown by~\cite{chesson:1990}, if
$S=R=\text{rank}(\mathsf{\Gamma})$, then exists a unique solution of the 
system $x_i+\sum_{j\ne i}\rho_{ij}x_j=K_i$, $i=1,\dots,S$. The resulting
equilibrium point will be interior if every species density remains strictly positive. 
Moreover,~\cite{chesson:1990} demonstrated that if there exists a unique interior 
equilibrium point for the dynamics~\eqref{eq:detdyn}, it will be globally 
stable if and only if $S=R=\text{rank}(\mathsf{\Gamma})$. Therefore, if competitive 
exclusion does not operate, a unique globally stable equilibrium point is reached 
and, conversely, if the Lotka-Volterra equations present an interior, globally stable 
equilibrium point, any positive initial condition will make the 
dynamics~\eqref{eq:detdyn} converge to the equilibrium point. This automatically 
ensures that none of the $S$ species is driven to extinction by competitive 
exclusion~\citep{macarthur:1970,takeuchi:1996}.

\section{Stability of the symmetric, deterministic model for $\rho\ge 1$}
\label{sec:appA}

In this Appendix we perform a stability analysis of the equilibrium points of the 
symmetric, deterministic dynamics in the competition regime $\rho\ge1$. As we 
mentioned before, when $\rho<1$ it can be shown that the interior equilibrium point is 
globally stable~\citep{hofbauer:1998,capitan:2015}, all boundary equilibria being
unstable.

We start by analyzing stability for $\rho=1$. In this case, the system is also stable 
and the initial condition determines the attractor which the dynamics converges to. Any 
equilibrium point is such that its densities satisfy
\begin{equation}
\sum_{j=1}^S x_j=K.
\end{equation}
Let $J(t)=\sum_{j=1}^S x_j(t)$. Then, by summing up the equations of the 
system~\eqref{eq:dyn} for $\rho=1$ we get
\begin{equation}
\frac{dJ}{dt}=rJ\left(1-\frac{J}{K}\right),
\end{equation}
which can be integrated and yields
\begin{equation}
J(t)=\left[\frac{1}{K}+\left(\frac{1}{J_0}-\frac{1}{K}\right)e^{-rt}\right]^{-1},
\end{equation}
$J_0$ being the initial condition for $J(t)$, $J_0=J(0)=\sum_{j=1}^S x_j(0)$. Therefore,
the dynamics of each species turns out to be decoupled,
\begin{equation}\label{eq:dxJ}
\frac{dx_i}{dt}=rx_i\left(1-\frac{J(t)}{K}\right).
\end{equation}
The equilibrium point to which~\eqref{eq:dxJ} converges is determined by the initial
condition vector $\vv{x_0}=(x_1(0),\dots,x_S(0))$. For two distinct species $i$ and 
$j$,~\eqref{eq:dxJ} implies that $\dot{x}_i/\dot{x}_j=x_i/x_j$. Integration yields 
$x_i(t)/x_j(t)=x_i(0)/x_j(0)$, which means that the proportions of population densities are 
conserved along the dynamics, whose orbits are reduced to straight lines starting from
the initial condition $\vv{x_0}$ along the direction determined by the vector $\vv{x_0}$. 
Therefore, the final equilibrium point is given by the intersection of the hyperplane 
$\sum_{j=1}^S x_j=K$ and the line that links the initial condition point, $\vv{x_0}$, 
and the origin. Any of these (infinite) equilibrium points will be stable, provided that the 
initial densities satisfy $x_i(0)\ge 0$ for all $i$.

The case $\rho>1$ leads to competitive exclusion. We analyze the asymptotic stability
of the $2^S$ equilibrium points with positive or zero densities. Without loss of generality,
for $0\le n\le S$ any equilibrium point will be of the form
\begin{equation}\label{eq:bound}
\vv{x}_n=\begin{pmatrix}
\hat{x}_n\vv{u}_{S-n}\\
\vv{0}_n
\end{pmatrix},
\end{equation}
where subscript $n$ indicates that $n$ densities are strictly equal to $0$, and 
$\vv{u}_n=(1,\dots,1)^{\text{T}}$ is a vector with $n$ entries equal to $1$. Any 
equilibrium with $n$ zero densities can be written as a permutation 
of~\eqref{eq:bound}, without altering the subsequent stability analysis. In addition,
the non-zero entries of $\vv{x}_n$ are the solutions of the system
\begin{equation}
(1-\rho)x_i+\rho\sum_{j=1}^{S-n} x_j=K,\qquad i=1,\dots,S-n.
\end{equation}
This system admits a single solution for which the $S-n$ species have equal densities,
$x_i=\hat{x}_n$, where
\begin{equation}
\hat{x}_n=\frac{K}{1-\rho+\rho(S-n)}.
\end{equation}

Our demonstration reduces to show that, if $\rho>1$, all the non-trivial equilibria act as 
repellors except for $n=1$, i.e., when only a single species survives. To this purpose we 
evaluate the eigenvalue spectra of the Jacobian matrix of the system~\eqref{eq:dyn}. 
The Jacobian matrix $\mathsf{J}$ can be expressed in a block form as
\begin{equation}
\mathsf{J}=\left(\begin{array}{c|c}
\mathsf{J}_{11} & \mathsf{J}_{12}\\
\hline
\mathsf{J}_{21} & \mathsf{J}_{22}\\
\end{array}\right),
\end{equation}
where
\begin{eqnarray}
&\displaystyle\mathsf{J}_{11}=-\frac{r\hat{x}_n}{K}\left[(1-\rho)\mathsf{I}_{S-n}+
\rho\vv{u}_{S-n}\vv{u}_{S-n}^{\text{T}}\right],\\
&\displaystyle\mathsf{J}_{12}=-\frac{r\rho\hat{x}_n}{K}\vv{u}_{S-n}\vv{u}_{n}^{\text{T}},\\
&\displaystyle\mathsf{J}_{21}=\vv{0}_n\vv{0}_{S-n}^{\text{T}},\\
&\displaystyle\mathsf{J}_{22}=\frac{r(1-\rho)\hat{x}_n}{K}\mathsf{I}_n,
\end{eqnarray}
and $\mathsf{I}_n$ is the $n\times n$ identity matrix and $\vv{0}_n=(0,\dots,0)^{\text{T}}$
is the zero vector with $n$ entries. Without loss of generality, eigenvectors can be written
as
\begin{equation}\label{eq:boundv}
\vv{v}=\begin{pmatrix}
\vv{a}_{S-n}\\
\vv{b}_n
\end{pmatrix},
\end{equation}
where $\vv{a}_{S-n}$ and $\vv{b}_n$ are column vectors with dimensions $S-n$ and $n$,
respectively. First let us assume that $\vv{b}_n=\vv{0}_n$. The spectral problem reduces to
\begin{equation}
\mathsf{J}_{11}\vv{a}_{S-n}=\lambda\vv{a}_{S-n}.
\end{equation}
Two different solutions arise: if $\vv{u}_{S-n}^{\text{T}}\vv{a}_{S-n}=0$, i.e., $\vv{a}_{S-n}$ is
orthogonal to $\vv{u}_{S-n}$, then the eigenvalue is
\begin{equation}\label{eq:l1}
\lambda=-\frac{r(1-\rho)\hat{x}_n}{K}
\end{equation}
with algebraic multiplicity $S-n-1$. However, if $\vv{a}_{S-n}\propto\vv{u}_{S-n}$
we find $\lambda=-r$ as an eigenvalue with algebraic multiplicity equal to $1$.

On the other hand, if $\vv{b}_n\ne\vv{0}_n$ we have to solve
\begin{equation}\label{eq:spe}
\begin{aligned}
&\mathsf{J}_{11}\vv{a}_{S-n}+\mathsf{J}_{12}\vv{b}_n=\lambda\vv{a}_{S-n},\\
&\mathsf{J}_{22}\vv{b}_n=\lambda\vv{b}_{n}.
\end{aligned}
\end{equation}
Since $\mathsf{J}_{22}$ is proportional to $\mathsf{I}_n$, we find 
\begin{equation}\label{eq:l3}
\lambda=\frac{r(1-\rho)\hat{x}_n}{K},
\end{equation}
with $n$ eigenvectors $\vv{b}_n=\vv{e}_n^{(i)}$ ($i=1,\dots,n$), $\vv{e}_n^{(i)}$ being the $i$-th 
vector of the canonical basis of $\mathbb{R}^n$. The algebraic multiplicity associated 
to~\eqref{eq:l3} is equal to $n$. Substituting these results into the first equation 
of~\eqref{eq:spe} yields
\begin{equation}
\left[(1-\rho)\mathsf{I}_{S-n}+\rho\vv{u}_{S-n}\vv{u}_{S-n}^{\text{T}}\right]\vv{a}_{S-n}+
\rho\vv{u}_{S-n}=-(1-\rho)\vv{a}_{S-n}.
\end{equation}
Given the structure of the matrices involved, we look for a solution of the form 
$\vv{a}_{S-n}=\alpha\vv{u}_{S-n}$, which implies a non-trivial value 
$\alpha=-\rho/[2(1-\rho)+\rho S]$.

Two eigenvalues determine the asymptotic stability of all the equilibrium points: 
$\lambda_1=-r(1-\rho)\hat{x}_n/K$ and $\lambda_2=-\lambda_1=r(1-\rho)\hat{x}_n/K$ 
(the third eigenvalue, $\lambda_3=-r$, is always negative). If $\rho>1$ and $0\le n< S-1$,
$\lambda_1>0$ and remains as an eigenvalue ---recall that its multiplicity is $S-n-1>0$. 
Therefore, any equilibrium with less than $S-1$ extinct species is asymptotically unstable 
---including the interior coexistence equilibrium. However, when only one species
survives ($n=S-1$), $\lambda_1$ is not an eigenvalue anymore and the two other 
eigenvalues remain: $\lambda_2=r(1-\rho)\hat{x}_n/K<0$ (with multiplicity $S-1$)
and $\lambda_3=-r<0$ (with multiplicity $1$). Thus, only the $S$ 
boundary equilibria with a single extant species are asymptotically stable.

Since the trivial equilibrium point (associated to complete extinction) is obviously 
unstable, we deduce that any orbit starting from an interior initial condition will be repelled
if it gets close to any equilibrium, except when the equilibrium point is formed by a 
single extant species. If the orbit enters the basin of attraction of any of those $S$
equilibria, it will end up in it asymptotically. This implies the extinction at a time of 
$S-1$ species if $\rho>1$.

\section{Numerical calculation of the steady-state probability distribution}
\label{sec:appB}

In order to compute numerically
the stationary joint distribution, we limit the infinite configuration space to
the set $\Xi\equiv\{0,1,\dots,n_{\mathrm{max}}\}^S$ by choosing 
$n_{\mathrm{max}}$ large enough so that the probability of finding a 
population number equal to $n_{\mathrm{max}}$ is negligible (we choose 
$n_{\mathrm{max}}$ as the integer part of $2K$, which fulfills the requirement).

The stationary distribution is obtained by solving the embedded Markov 
chain associated to the continuous-time Markov process~\citep{karlin:1975}.
The transition matrix of the embedded Markov chain is defined
by the transition probabilities
\begin{equation}
\text{Pr}\{\vv{n}\to\vv{n}\pm\vv{e}_i\}=\frac{q_i^{\pm}(\vv{n})}{\Lambda(\vv{n})},
\end{equation}
where $\Lambda(\vv{n})=\sum_{i=1}^S[q_i^{+}(\vv{n})+
q_i^{-}(\vv{n})]$. The remaining transitions have zero probability.
Elementary events (overall births and deaths) take place after exponential
times, so that the time lapsed to the next event is drawn from a random
variable $\tau$ with cumulative distribution 
$\text{Pr}(\tau\le t)=1-e^{-\Lambda(\vv{n})t}$. Once the steady-state
distribution $\boldsymbol{\varphi}=(\varphi(\vv{n}))$ of the embedded 
Markov chain ---i.e., the left-eigenvector of the transition matrix with eigenvalue 
1--- has been determined, according to the mean time spent by the process at 
state $\vv{n}$, the probability of finding the continuous-time Markov 
process at state $\vv{n}$ turns out to be~\citep{cinlar:1975}
\begin{equation}\label{eq:P}
P(\vv{n})=\frac{\varphi(\vv{n})\Lambda(\vv{n})^{-1}}{\sum_{\boldsymbol{m}\in\Xi}
\varphi(\vv{m})\Lambda(\vv{m})^{-1}}.
\end{equation}

\section{Coefficient of variation of population abundances in the limit of large carrying capacity}
\label{sec:appC}

In this section we derive an analytical expression for the coefficient of variation of population
abundances in the small stochasticity limit. In the case of demographic stochasticity, low
variability levels can be obtained for large population sizes or, equivalently, in the limit of 
large carrying capacity. We build on the Gaussian approximation for the joint probability 
distribution, which is valid in the limit $K\gg 1$ since the probability of extinction configurations
is expected to be negligible, and can be fully calculated for a generic community of size $S$.

The Gaussian approximation can be obtained as the solution to the Fokker-Planck equation 
deduced from the master equation~\eqref{eq:master}. We do not reproduce its derivation
here; it can be found at the Supplemental Information of~\cite{capitan:2015}. 
Under this approximation, the joint probability distribution is expressed as
\begin{equation}
\Pi(\vv{n})=\frac{1}{Z}\text{exp}\{-(\vv{n}-\hat{x}\vv{u}_S)^{\text{T}}\mathsf{Q}(\vv{n}-\hat{x}\vv{u}_S)\},
\end{equation}
where $Z$ is an appropriate normalization factor, 
\begin{equation}
\hat{x}=\frac{K}{2(1-\rho+\rho S)}\left[1+\sqrt{1+\frac{4\mu(1-\rho+\rho S)}{rK}}\right],
\end{equation}
and the covariance matrix $\mathsf{Q}^{-1}$ is given by
\begin{equation}\label{eq:cov}
\mathsf{Q}^{-1}=\frac{b}{2a}\left(\mathsf{I}_S-\frac{c}{a+cS}\vv{u}_S\vv{u}_S^{\text{T}}\right).
\end{equation}
In terms of model parameters, $a=r\hat{x}(1-\rho)/K+\mu/\hat{x}$, $b=2(r^+\hat{x}+\mu)$, and 
$c=r\rho\hat{x}/K$. In the large carrying capacity limit, the average population abundance is 
expressed through a series expansion on powers of $K$ as
\begin{equation}
\langle n\rangle=\hat{x}=\frac{K}{1-\rho+\rho S}+\frac{\mu}{r}+\mathcal{O}(K^{-1}).
\end{equation}
Similarly, series expansions give
\begin{equation}
\begin{aligned}
&a=\frac{r(1-\rho)}{1-\rho+\rho S}+\mathcal{O}(K^{-1}),\\
&b=\frac{2r^+K}{1-\rho+\rho S}+2\mu\left(\frac{r^+}{r}+1\right)+\mathcal{O}(K^{-1}),\\
&c=\frac{r\rho}{1-\rho+\rho S}+\mathcal{O}(K^{-1}).
\end{aligned}
\end{equation}
Inserting these expressions into~\eqref{eq:cov} yields, up to order $K^0$, the approximation
\begin{equation}\label{eq:cov2}
\mathsf{Q}^{-1}=\frac{r^+K}{r(1-\rho)}\left(\mathsf{I}_S-
\frac{\rho}{1-\rho+\rho S}\vv{u}_S\vv{u}_S^{\text{T}}\right)+\mathcal{O}(K^0).
\end{equation}
Therefore the standard deviation of population abundance, $\sigma_n$, can be obtained 
as the square root of diagonal elements of matrix $\mathsf{Q}^{-1}$, 
\begin{equation}\label{eq:std}
\sigma_n=\sqrt{\frac{r^+(1-2\rho+\rho S)}{r(1-\rho)(1-\rho+\rho S)}}K^{1/2}+\mathcal{O}(K^{-1/2}),
\end{equation}
which (not surprisingly) scales with $K$ as $K^{1/2}$. Finally, the coefficient of variation of 
population abundances is expressed, in the limit $K\gg 1$, as
\begin{equation}\label{eq:CV}
\nu=\frac{\sigma_n}{\langle n\rangle}=\sqrt{\frac{r^+(1-2\rho+\rho S)(1-\rho+\rho S)}{r(1-\rho)}}K^{-1/2}.
\end{equation}
Strictly speaking, the deterministic scenario ($\nu=0$) is only achieved in the limit $K\to\infty$. 
However, low stochasticity regimes can be assessed using Eq.~\eqref{eq:CV}: if the actual 
coefficient of variation is close to that yielded by the Gaussian approximation, both of which 
are small for large $K$, then extinction configurations are precluded and the variability of 
populations with respect to the mean value is small. We adopt, as a practical definition for 
low stochasticity, the parameter combinations for which the actual coefficient of variation is 
close to the approximation given by Eq.~\eqref{eq:CV}.

\section{Stability analysis for two deterministic models with non-symmetric competition}
\label{sec:appD}

Here we analyze the stability of the equilibrium points of the two non-symmetric, three-species 
competitive dynamics of the form~\eqref{eq:nosym} considered in the main text, for which 
the interaction matrix is written as
\begin{equation}\label{eq:rhons}
\mathsf{R}=(\rho_{ij})=
\begin{pmatrix}
1 & \rho & \rho+\delta_1\\
\rho & 1 & \rho+\delta_2\\
\rho+\delta_1 & \rho+\delta_2 & 1
\end{pmatrix},
\end{equation}
$\delta_1$ and $\delta_2$ being two positive numbers. The sign of equilibrium densities 
and the corresponding eigenvalues of the Jacobian matrix determine the ranges of $\rho$
for which the system is stable. We impose the condition $\rho\ge 0$ for all interaction 
coefficients to remain positive. Since the growth rate $r>0$, in both cases the full extinction 
equilibrium $\vv{\hat{x}}=(0,0,0)$ is unstable.

In the first example, $\delta_1=0.1$ and $\delta_2=0.05$. Species carrying capacities are 
non-uniform: $K_1=40$, $K_2=16$, and $K_3=20$. Although the expressions are too
cumbersome to be reproduced here, it can be shown that the interior equilibrium point
$\vv{\hat{x}}=\mathsf{R}^{-1}\vv{K}$, with $\vv{K}=(K_1,K_2,K_3)^{\mathrm{T}}$, 
has three positive densities and is asymptotically stable if and only if $0\le\rho<0.4$.

We now consider all boundary equilibria. In what follows the eigenvalues $\lambda$ 
of the stability (Jacobian) matrix are expressed as $\lambda=\lambda' r$, i.e., they 
are scaled by the growth rate $r>0$:
\begin{enumerate}[label=(\alph*)]
\item The first and second coordinates of
\begin{equation}\vv{\hat{x}}=\left(\frac{8(5-2\rho)}{1-\rho^2},\frac{8(2-5\rho)}{1-\rho^2},0\right),
\end{equation}
are positive if and only if $0\le\rho<0.4$ or $\rho>2.5$. The scaled eigenvalues $\lambda'$
are given by 
\begin{equation}
\lambda'\in\left\{-1,\frac{38-131\rho+90\rho^2}{50(1-\rho^2)},-\frac{(5-2\rho)(2-5\rho)}{10(1-\rho^2)}\right\}.
\end{equation}
The condition $\lambda'_i<0$ for $i=1,2,3$ yields $1.055<\rho<2.5$. Therefore, this 
point is never feasible and stable at the same time.
\item For
\begin{equation}
\vv{\hat{x}}=\left(\frac{200(19-10\rho)}{99-20\rho-100\rho^2},0,
\frac{800(2-5\rho)}{99-20\rho-100\rho^2}\right),
\end{equation}
we require positivity for the first and third entries, which yields $0\le\rho<0.4$ or $\rho>1.9$. 
On the other hand, the eigenvalues in this case are 
\begin{equation}
\lambda'\in\left\{-1,\frac{94-345\rho-275\rho^2}{(9-10\rho)(11+10\rho)},
-\frac{2(2-5\rho)(19-10\rho)}{(9-10\rho)(11+10\rho)}\right\}.
\end{equation}
Stability implies $0.9<\rho<1.9$, which is incompatible with the feasibility condition.
\item The third equilibrium with a single extinct species is
\begin{equation}
\vv{\hat{x}}=\left(0,\frac{2000(3-4\rho)}{399-40\rho-400\rho^2},
\frac{1280(6-5\rho)}{399-40\rho-400\rho^2}\right).
\end{equation}
This point is feasible if and only if $0\le\rho<0.75$ or $\rho>1.2$. The (scaled) eigenvalues of the
Jacobian matrix are 
\begin{equation}
\lambda'\in\left\{-1,\frac{-20(3-4\rho)(6-5\rho)}{(19-20\rho)(21+20\rho)},
\frac{1899-1830\rho-200\rho^2}{(19-20\rho)(21+20\rho)}\right\}.
\end{equation}
Since the system of inequalities $\lambda'_i<0$ ($i=1,2,3$) turns out to be 
incompatible, this point is unstable for all values of $\rho$.
\item $\vv{\hat{x}}=(40,0,0)$: the scaled eigenvalues are $\{-1,(2-5\rho)/2,2(2-5\rho)/5\}$. 
This point is stable for $\rho>0.4$.
\item $\vv{\hat{x}}=(0,16,0)$: $\lambda'\in\{-1,(5-2\rho)/5,4(6-5\rho)/25\}$. The 
equilibrium point is stable if and only if $\rho>2.5$.
\item $\vv{\hat{x}}=(0,0,20)$: $\lambda'\in\{-1,(19-10\rho)/20,5(3-4\rho)/16\}$. 
Asymptotic stability is achieved for $\rho>1.9$.
\end{enumerate}
As a result, in the range $0\le\rho<0.4$, the only stable point is the coexistence
equilibrium. However, for $\rho>0.4$, only two-extinct species equilibria remain
asymptotically stable. Since the eigenvalues are continuous functions of model parameters,
close to this example we can find multiple non-symmetric systems that exhibit a grouped, 
two-species extinction as $\rho$ increases. 

The second example shows that multiple stable equilibria can co-occur when interactions
are chosen non-symmetrically. In this case, we have taken $\delta_1=0.4$, $\delta_2=0.3$
and $K_1=K_2=K_3=15$. The densities of the interior equilibrium point are expressed as
\begin{equation}\label{eq:inteq}
\begin{aligned}
&x_1=15(9-10\rho)(7-10\rho)/D(\rho),\\
&x_2=30(3-5\rho)(11-10\rho)/D(\rho),\\
&x_3=150(3-10\rho)(1-\rho)/D(\rho),
\end{aligned}
\end{equation}
where $D(\rho)=75-116\rho-160\rho^2+200\rho^3$. The 
feasibility analysis of the equilibrium point yields, for $\rho\ge 0$, the ranges
$0\le\rho<0.3$ or $0.7<\rho<0.9$ or $\rho>1.1$. The eigenvalues of the stability 
matrix can be fully calculated, although their expressions are too cumbersome 
to be reproduced here. It is easy to check that the three eigenvalues are negative 
if and only if $0\le\rho<0.3$. Therefore, this equilibrium point is interior and
asymptotically stable if and only if $0\le\rho<0.3$.

We now summarize the stability analysis for boundary equilibria. All of them
are feasible, so stability is only conditioned by the sign of eigenvalues (which
are all real):
\begin{enumerate}[label=(\alph*)]
\item $\vv{\hat{x}}=\left(\frac{15}{1+\rho},\frac{15}{1+\rho},0\right)$: the scaled 
eigenvalues $\lambda'$ are $\left\{-1,\frac{3-10\rho}{10(1+\rho)},
\frac{-1+\rho}{1+\rho}\right\}$. Therefore, this point is stable for $0.3<\rho<1$.
\item $\vv{\hat{x}}=\left(\frac{75}{7+5\rho},0,\frac{75}{7+5\rho}\right)$: 
$\lambda'\in\left\{-1,\frac{11-10\rho}{2(7+5\rho)},\frac{-3+5\rho}{7+5\rho}\right\}$. 
The stability conditions form an unfeasible problem, so this point turns out to be 
unstable for any $\rho$.
\item $\vv{\hat{x}}=\left(0,\frac{150}{13+10\rho},\frac{150}{13+10\rho}\right)$: 
$\lambda'\in\left\{-1,\frac{9-10\rho}{13+10\rho},\frac{-7+10\rho}{13+10\rho}\right\}$.
Again, this point is unstable for all values of competitive overlap.
\item $\vv{\hat{x}}=(15,0,0)$: the scaled eigenvalues are 
$\left\{-1,\frac{3}{5}-\rho,1-\rho\right\}$. This point is stable for $\rho>1$.
\item $\vv{\hat{x}}=(0,15,0)$: $\lambda'\in\left\{-1,\frac{7-10\rho}{10},1-\rho\right\}$. 
The equilibrium point is stable if and only if  $\rho>1$.
\item $\vv{\hat{x}}=(0,0,15)$: $\lambda'\in\left\{-1,\frac{3-5\rho}{5},\frac{7-10\rho}{10}\right\}$. 
Stability is attained for $\rho>0.7$.
\end{enumerate}
Consequently, in the range $0.3<\rho<0.7$ the only stable 
equilibrium point is $\left(\frac{15}{1+\rho},\frac{15}{1+\rho},0\right)$. However, for 
$0.7<\rho<1$ two stable equilibria co-occur: the former and a two-extinct 
species equilibrium, $\left(0,0,15\right)$. Depending on initial conditions, the 
dynamics can lead to one of them or to the other. For $\rho>1$, however,
the three equilibria with a single extant species are the only ones that remain 
asymptotically stable.

This example shows how the cascade of extinctions in non-symmetric, 
deterministic models can be far from being determined due to the co-occurrence 
of multiple stable equilibria.

\bibliographystyle{model2-names}
\bibliography{ecology}







\end{document}